\documentclass[10pt,francais]{llncs}
\usepackage[a4paper]{geometry}
\usepackage[T1]{fontenc}
\usepackage[latin1]{inputenc}
\usepackage{aeguill}

\usepackage[frenchb]{babel} % nous \'erivons du fran\c{c}ais ou de l'anglais
\usepackage{multirow}
\usepackage{mathrsfs}
\usepackage{color}
\usepackage{theorem}
\usepackage[french,vlined,ruled]{algorithm2e}
\usepackage{amssymb}

\newtheorem{rem}{Remarque}

\newtheorem{defs}{D\'efinition}

\newenvironment{engabstract}{%
      \list{}{\advance\topsep by0.35cm\relax\small
      \leftmargin=1cm
      \labelwidth=0pt
      \listparindent=0pt
      \itemindent\listparindent
      \rightmargin\leftmargin}\item[\hskip\labelsep
                                    \bfseries Abstract]}
    {\endlist}

\title{RAPPORT DE RECHERCHE LIPN\\
~\\
Le probl\`eme du plus court chemin contraint}
\author{Olivier Laval\inst{1} \and Sophie Toulouse\inst{1} \and Anass Nagih\inst{2}}
\institute{LIPN, Universit\'e de Paris-Nord\\ 99 avenue Jean-Baptiste Cl\'ement 93430 Villetaneuse, France\\ \email{\{olivier.laval,sophie.toulouse\}@lipn.univ-paris13.fr}
\and LITA, Universit\'e Paul Verlaine\\ Ile du Saulcy 57045 Metz Cedex 01, France\\
\email{anass.nagih@univ-metz.fr}
}

\begin{document}
\date{21 d\'ecembre 2006}
\thispagestyle{plain}
\maketitle
\begin{center}
26 d\'ecembre 2006
\end{center}

\begin{abstract}
Cet article propose un tour d'horizon des m\'ethodes approch\'ees et exactes, de leur performance et de leur complexit\'e th\'eorique, pour diff\'erentes versions du probl\`eme de plus court chemin. L'\'etude propos\'ee est faite dans l'optique d'am\'eliorer la r\'esolution d'un probl\`eme plus g\'en\'eral de couverture dans le cadre d'un sch\'ema de g\'en\'eration de colonnes, dont le plus court chemin appara\^it comme le sous-probl\`eme. 
\end{abstract}

\begin{engabstract}
This article provides an overview of the performance and the theoretical complexity of approximate and exact methods for various versions of the shortest path problem. The proposed study aims to improve the resolution of a more general covering problem within a column generation scheme in which the shortest path problem is the sub-problem.
\end{engabstract}

\section{Introduction}\label{sec-intro}
Un probl\`eme des plus courants en optimisation combinatoire est celui de la recherche de plus courts chemins dans un graphe. Ce probl\`eme se pr\'esente comme suit~: \'etant donn\'e un graphe et une fonction co\^ut sur les arcs, le probl\`eme consiste \`a trouver le chemin le moins co\^uteux d'un sommet choisi \`a un autre. Il se r\'esoud ais\'ement gr\^ace \`a de nombreux algorithmes polynomiaux  (Bellman \cite{Bellman1958}, Dijkstra \cite{Dijkstra1959}, ...). N\'eanmoins, l'ajout de contraintes sur le chemin (essentiellement, des contraintes de type ``sac-\`a-dos'') le rend plus difficile \`a r\'esoudre. Ce rapport de recherche pr\'esente donc diff\'erents algorithmes exacts ou approch\'es pour r\'esoudre le probl\`eme du Plus Court Chemin Contraint (not\'e PCCC par la suite). Ce travail se place dans le cadre plus g\'en\'eral de r\'esolution d'un probl\`eme de couverture de t\^aches par des v\'ehicules qui doivent respecter certaines contraintes dont le PCCC est, dans une d\'ecomposition classique  de type Dantzig Wolfe, le sous-probl\`eme. Aussi, les r\'esultats propos\'es seront le plus souvent compar\'es dans le cadre du d\'eroulement de tels sch\'emas. 
Ce document est organis\'e comme suit~: la deuxi\`eme section pr\'esentera le probl\`eme PCCC de mani\`ere tr\`es g\'en\'erale, avec ses diff\'erentes variantes. Plusieurs m\'ethodes de r\'eduction des instances, qui constituent une \'etape pr\'eliminaire \`a la r\'esolution du probl\`eme, sont expos\'ees dans la troisi\`eme section. La r\'esolution exacte sera abord\'ee dans la quatri\`eme section. Enfin, puisque le probl\`eme est $\mathbf{NP}$-difficile, la cinqui\`eme section portera sur sa r\'esolution approch\'ee.

\section{Description du probl\`eme}\label{sec-desc}
\subsection{Formalisation}\label{sub-desc-form} 
Soit $G=(V,A)$ un graphe orient\'e o\`u $V$ est l'ensemble des sommets ($|V| = n$) et $A \subseteq V \times V$ l'ensemble des arcs ($|A| = m$) et soit $\mathscr{R}$ un ensemble de ressources ($|\mathscr{R}| =R$). \`A chaque arc $(i,j)\in A$ sont associ\'es un co\^ut $c_{ij}$ et un vecteur de consommation de ressource $(t^r_{ij})_{r=1,\ldots,R}$ positif ($\forall r=1,\ldots,R,\ t^r_{ij} \geq 0$). Le graphe ne doit pas comporter de cycle absorbant\footnote{Un cycle absorbant est un cycle dont la somme des valuations des arcs est n\'egative.}.\\

Un chemin $P_{x \to y}$ entre deux sommets $x$ et $y$ est une s\'equence d'arcs : 
$$\displaystyle P_{x \to y} = \bigcup_{i=1}^{p}\{(u_i,v_i)\}\ \mathrm{tel\ que}\ (u_i,v_i) \in A,\ u_{i+1} = v_i,\ u_1 = x\ \mathrm{et}\ v_p = y$$
La longueur de ce chemin est alors $p$. Un chemin $P_{x \to y}$ allant du sommet $x$ au sommet $y$ a un co\^ut dont la formule est $\displaystyle C(P_{x \to y}) = \sum_{(u,v) \in P_{x \to y}} c_{uv}$.\\
Soient $s$ et $t$ deux sommets distincts privil\'egi\'es du graphe, appel\'es respectivement source et puits. Le probl\`eme du Plus Court Chemin Contraint (PCCC) consiste alors \`a trouver un chemin de co\^ut minimal entre la source et le puits satisfaisant certaines contraintes de ressource (ces contraintes seront d\'etaill\'ees par la suite). Si l'ensemble $\mathscr{R}$ des ressources est vide ($R=0$), on est ramen\'e au probl\`eme usuel de plus court chemin qui est polynomial (\cite{Bellman1958}, \cite{Dijkstra1959}). En revanche, la consid\'eration d'une seule ressource rend d\'ej\`a le probl\`eme d'optimisation $\mathbf{NP}$-dur, et ce m\^eme lorsque les co\^uts et les consommations de ressource sont suppos\'es \^{e}tre entiers positifs \cite{Garey1979} \cite{Dror1994}. Enfin, le probl\`eme consistant \`a d\'ecider seulement s'il existe ou non un chemin r\'ealisable est $\mathbf{NP}$-complet d\`es lors que l'on consid\`ere deux ressources ou plus.

\subsection{Contraintes de ressource}\label{sub-desc-contrainte}
Les contraintes de ressource s'expriment \`a l'aide d'un vecteur consommation de ressource $T$ d\'ependant du chemin consid\'er\'e dont chaque coordonn\'ee repr\'esente une ressource du probl\`eme.

Deux types de contraintes de ressource sont usuellement consid\'er\'es. Tout d'abord, les contraintes de ressource dites {\em finales}, o\`u la somme sur tous les arcs du chemin de la source au puits des quantit\'es de ressource consomm\'ees doit entrer dans une fen\^etre d\'efinie au puits~; ensuite, les contraintes de ressource dites {\em \`a fen\^etres de temps} o\`u, \`a chaque sommet $i$, sont associ\'ees $R$ fen\^etres de ressource $[a_i^r,b_i^r]$, $r=1,\ldots,R$, r\'eduisant l'intervalle des valeurs possibles pour la quantit\'e de ressource~$r$ pouvant \^etre utilis\'ee avant d'atteindre le sommet~$i$. Pour ce type de contrainte, il existe deux m\'ethodes de calcul du vecteur consommation de ressource sur un chemin de la source \`a un sommet :
\begin{itemize}
\item sans attente permise : 
		$$\forall (P_{s \to j} = P_{s \to i} \cup (i,j)),\ T^r(P_{s \to j}) = T^r(P_{s \to i}) + t_{ij}^r$$
\item avec attente : 
		$$\forall (P_{s \to j} = P_{s \to i} \cup (i,j)),\ T^r(P_{s \to j}) = \max \{a_j^r, T^r(P_{s \to i}) + t_{ij}^r\}$$
\end{itemize}
Un chemin de la source au puits est dit r\'ealisable si, en chacun de ses sommets, le vecteur consommation de ressource sur le sous-chemin de $s$ \`a ce sommet est dans la fen\^etre de ressource de ce sommet~; formellement~:
$$\begin{array}{ll}
                    &\textrm{le chemin }P_{s \to t}\textrm{ est r\'ealisable}\\
\textit{ssi} 	&\forall j \in P_{s \to t},\ \forall r=1,\ldots,R,\ a_j^r \leq T^r(P_{s \to j}) \leq b_j^r
\end{array}$$

De cette in\'egalit\'e d\'ecoule deux propri\'et\'es g\'en\'erales :
\begin{itemize}
 \item dans les deux m\'ethodes (avec ou sans attente) :
$$\forall (i,j)\in A,\ \forall r=1,\ldots,R,\ a_i^r + t_{ij}^r \leq b_j^r$$
 \item dans la m\'ethode sans attente permise :
$$\forall (i,j)\in A,\ \forall r=1,\ldots,R,\ b_i^r + t_{ij}^r \leq a_j^r$$
\end{itemize}
Les arcs ne v\'erifiant pas ces in\'egalit\'es peuvent \^etre directement supprim\'es car ils n'appartiennent \`a aucune solution r\'ealisable du probl\`eme.

Le premier type de contrainte se ram\`ene ais\'ement au second en attribuant \`a tous les sommets du graphe et pour chaque ressource la fen\^etre $[0,b_t^r]$, o\`u  $b_t^r$ est le majorant de consommation d\'efinie au puits pour cette ressource. Toutes les instances consid\'er\'ees dans ce document seront donc, sans mention expresse du contraire, du second type. 

\subsection{Relation avec la g\'en\'eration de colonnes}\label{sub-desc-gc} %% MODIF_ST
Dans certaines mod\'elisations de probl\`emes lin\'eaires en nombres entiers dites \`a formulation \og chemins \fg{}, les variables du probl\`eme repr\'esentent des chemins du graphe. Le nombre de chemins dans un graphe \'etant potentiellement exponentiel, il en est de m\^eme du nombre de variables. L'utilisation d'un algorithme de r\'esolution (simplexe par exemple) ne peut \^etre envisag\'ee pour ce mod\`ele puisque ce dernier ne peut \^etre explicit\'e. En revanche, un probl\`eme avec un sous-ensemble de variables de taille raisonnable, appel\'e probl\`eme ma\^itre restreint (PMR), peut \^etre r\'esolu~; la r\'esolution de ce dernier permet de calculer les co\^uts r\'eduits associ\'es \`a chaque arc du graphe, que l'on peut interpr\'eter comme le co\^ut d'opportunit\'e \`a emprunter un arc donn\'e (par le biais d'un chemin utilisant cet arc). La variable pouvant \^etre ajout\'ee au PMR est alors trouv\'ee par r\'esolution du sous-probl\`eme qui consiste en la d\'etermination d'un plus court chemin contraint pour le crit\`ere de co\^ut r\'eduit. Par exemple, dans le cas d'une minimisation, si le co\^ut r\'eduit total d'un chemin est n\'egatif, il est potentiellement am\'eliorant pour le probl\`eme ma\^itre et constitue ainsi un bon candidat \`a \^etre int\'egr\'e au PMR. Dans cette approche appel\'ee {\em g\'en\'eration de colonnes}, l'optimalit\'e est atteinte lorsqu'il n'existe plus de chemin am\'eliorant (chemin de co\^ut r\'eduit n\'egatif si l'on consid\`ere la r\'esolution continue du probl\`eme ma\^{\i}tre). Notons que, pour la convergence, il n'est pas n\'ecessaire de trouver un chemin optimal : un chemin am\'eliorant suffit. De plus, \`a chaque it\'eration, plusieurs chemins am\'eliorants peuvent venir enrichir le PMR. \\
Le probl\`eme de plus court chemin contraint \'etant $\mathbf{NP}$-difficile, le temps de r\'esolution des sous-probl\`emes n'est pas ma\^itris\'e~; d'o\`u l'int\'er\^et d'utiliser des algorithmes d'approximation, du moins en d\'ebut de sch\'ema, nous permettant de trouver rapidement des chemins am\'eliorants (la r\'solution exacte demeurant in\'evitable en fin de sch\'ema pour prouver l'optimalit\'e).

\section{R\'eduction du probl\`eme avant sa r\'esolution}\label{sec-reduc}
Quelques travaux pr\'ec\'edant la r\'esolution peuvent \^etre effectu\'es dans le but de r\'eduire le graphe (suppression de sommets ou d'arcs, r\'eduction de l'amplitude des fen\^etres de temps). En outre, ces traitements permettent parfois de d\'etecter les instances non r\'ealisables ({\em i.e.}, instances sur lesquelles tout chemin viole les contraintes de ressource) et d'exhiber un majorant et un minorant.

\subsection{R\'eduction des fen\^etres de temps}\label{sub-reduc-reduc}
Les deux types de contraintes de ressource forment en fait une seule classe de probl\`emes o\`u chaque sommet poss\`ede des fen\^etres de temps. En chaque sommet, les bornes des fen\^etres de temps pour chaque ressource doivent v\'erifier quelques relations d\'ependant de ses pr\'ed\'ecesseurs et de ses successeurs. L'ensemble des pr\'ed\'ecesseurs de $i$ se notera $pred(i)$, l'ensemble des ses successeurs $succ(i)$. Ainsi, pour une ressource donn\'ee $r$, le minorant d'une fen\^etre du sommet $j$ ne doit pas \^etre plus petit que le minorant d'un pr\'ed\'ecesseur $i$ du sommet auquel le temps de trajet entre les sommets $i$ et $j$ est ajout\'e. Cela donne l'\'equation suivante~:
$$
a^r_j = \max \{ a^r_j, \min_{i \in pred(j)} \{ a^r_i + t^r_{ij}\} \}
$$
Une \'equation similaire est valable pour les majorants~:
$$
b^r_j = \min \{ b^r_j, \max_{i \in pred(j)} \{ b^r_i + t^r_{ij}\} \}
$$
Maintenant, en consid\'erant les successeurs, le minorant d'un sommet $j$ ne doit pas \^etre plus petit que le minorant d'un successeur $i$ du sommet auquel le temps de trajet de $j$ \`a $i$ est soustrait. Cela donne l'\'equation suivante~:
$$
a^r_j = \max \{ a^r_j, \min_{i \in succ(j)} \{ a^r_i - t^r_{ji}\} \}
$$
Une nouvelle \'equation similaire est valable pour les majorants~:
$$
b^r_j = \min \{ b^r_j, \max_{i \in succ(j)} \{ b^r_i - t^r_{ji}\} \}
$$
Si, pour un sommet $i$ et une ressource $r$, la condition $a^r_i > b^r_i$ est remplie, alors le sommet $i$ doit \^etre supprim\'e du graphe (car il est inaccessible), ainsi que tous ses arcs incidents. De m\^eme, dans le cas o\`u l'attente n'est pas permise, s'il existe un arc $(i,j)$ tel que $b^r_i + t^r_{ij} < a^r_j$, alors l'arc $(i,j)$ peut \^etre supprim\'e.\\

\begin{algorithm}\label{algo-reduction}
\caption{REDUCTION~: R\'eduction des fen\^etres de temps (graphe acyclique)}
\Entree{\\
$G = (V,A)$ et $\forall i \in V, \forall r=1,\ldots,R, [a_i^r;b_i^r]$ la fen\^etre de la ressource $r$ pour le sommet $i$~; les sommets sont num\'erot\'es dans l'ordre topologique}
\BlankLine
\Sortie{\\
$G'=(V',A')$ un sous-graphe partiel de $G$ avec les fen\^etres de ressource r\'eduites}
\BlankLine
{$V'\leftarrow V;~A'\leftarrow A$}\;
\Repeter{$recommencer = false$}{
  $recommencer \leftarrow false$\;
  \Pour{$i$ allant de $1$ \`a $n$}{
    \Pour{$r$ allant de $1$ \`a $R$}{
      $minp \leftarrow \infty;~maxp \leftarrow 0$\;
      \PourTous{$j \in pred(i)$}{
        $minp \leftarrow \min\{minp,a_j^r+t_{ji}^r\}$\;
        $maxp \leftarrow \max\{maxp,b_j^r+t_{ji}^r\}$\;
      }
      \lSi{$a_i^r < minp$}{$a_i^r \leftarrow minp$}\;
      \lSi{$b_i^r > maxp$}{$b_i^r \leftarrow maxp$}\;
    }
  }
  \Pour{$i$ allant de $n$ \`a $1$}{
    \Pour{$r$ allant de $1$ \`a $R$}{
      $mins \leftarrow \infty;~maxs \leftarrow 0$\;
      \PourTous{$j \in succ(i)$}{
        $mins \leftarrow \min\{mins,a_j^r-t_{ji}^r\}$\;
        $maxs \leftarrow \max\{maxs,b_j^r-t_{ji}^r\}$\;
      }
      \lSi{$a_i^r < mins$}{$a_i^r \leftarrow mins$}\;
      \lSi{$b_i^r > maxs$}{$b_i^r \leftarrow maxs$}\;
    }
  }
  \Si{attente non permise et $\exists (i,j) \in A',\exists r=1,\ldots,R, b_i^r + t_{ij}^r < a_j^r$}{
    $A' \leftarrow A' \setminus \{(i,j)\}$\;
    $recommencer \leftarrow true$\;
  }
  \Si{$\exists i \in V', \exists r=1,\ldots,R, b_i^r < a_i^r$}{
    $V' \leftarrow V'\setminus\{i\}$\;
    $A' \leftarrow A' \setminus \{(u,v) \in A', u = i\ ou\ v = i\}$\;

    $recommencer \leftarrow true$\;
  }
}
\end{algorithm}

L'algorithme~\ref{algo-reduction} v\'erifie donc les \'equations pr\'ec\'edentes pour chaque sommet. Dans le cas d'un graphe acyclique, il suffit de traiter les sommets dans l'ordre topologique pour les \'equations utilisant les pr\'edecesseurs, puis dans l'ordre inverse pour les \'equations concernant les successeurs. La proc\'edure recommence si et seulement si un sommet ou un arc peut \^etre supprim\'e. La complexit\'e de cette proc\'edure est donc en $\mathcal{O}(n^3\ R)$. Dans le cas cyclique, rien ne garantit qu'aucune fen\^etre ne pourra encore changer apr\`es l'application de la proc\'edure, et ce m\^eme si celle-ci n'induit pas de suppression de sommets ou d'arcs. La proc\'edure pr\'ec\'edente est donc r\'ep\'et\'ee tant qu'une borne change ou qu'un sommet ou un arc est \'elimin\'e. Pour cet algorithme, la complexit\'e en temps de la boucle \textit{r\'ep\'eter} est en $\mathcal{O}(n^2\ R)$. Une premi\`ere approximation na\"ive du nombre de fois o\`u cette boucle est effectu\'ee dans le cas de fen\^etres enti\`eres est $\displaystyle{\mathcal{O}(n\ \max_{i\in V} \{b_i - a_i \})}$ (r\'eduction d'au moins une unit\'e de la fen\^etre pour un sommet \`a chaque it\'eration). La complexit\'e totale est donc $\displaystyle{\mathcal{O}(n^3\ R\ \max_{i\in V} \{b_i - a_i \})}$.

\subsection{Pr\'etraitement pour contraintes de ressource finales}\label{sub-reduc-final}
Le but de cette proc\'edure de pr\'etraitement, pr\'esent\'ee dans \cite{Aneja1983} et am\'elior\'ee dans\cite{Dumitrescu2003} est, comme pr\'ec\'edemment, de r\'eduire le graphe mais en plus de fournir un minorant et un majorant pour la valeur du probl\`eme. Elle exploite le fait suivant~: trouver le plus court chemin non contraint entre deux sommets a la m\^eme complexit\'e (et prend aussi le m\^eme temps de calcul) que trouver le plus court chemin entre un sommet et {\em tous} les autres (ou {\em tous les autres} et {\em un} sommet). Les plus courts chemins suivant chaque m\'etrique (co\^ut ou ressource) sont donc calcul\'es entre la source et tous les sommets, ainsi qu'entre tous les sommets et le puits. Ensuite, en recombinant ces plus courts chemins, certains peu co\^uteux mais non r\'ealisables peuvent \^etre d\'etect\'es, ce qui permet d'am\'eliorer le minorant~; inversement,  des solutions r\'ealisables peuvent \^etre exhib\'ees, ce qui fournit un majorant au probl\`eme. Enfin, ce traitement permet d'\'elaguer le graphe en d\'eterminant des arcs et des sommets qui n'appartiennent \`a aucun chemin r\'ealisable ou qui n'appartiennent \`a aucun chemin optimal.\\

\noindent Le d\'eroulement de l'algorithme est le suivant~:
\begin{itemize}
 \item Tout d'abord, les chemins les moins co\^uteux de la source \`a tous les sommets sont calcul\'es. Cela permet \'eventuellement de d\'etecter des instances non r\'ealisables (absence de chemin de la source au puits) ou d'exhiber un chemin optimal (un plus court chemin de la source au puits v\'erifie les contraintes de ressource) ou encore, d'exhiber un chemin non r\'ealisable de co\^ut minimum permettant de mettre \`a jour le minorant pour le probl\`eme.
 \item Ensuite, pour chaque ressource, les chemins les moins consommateurs de cette ressource allant de la source \`a chaque sommet sont calcul\'es. Ils permettent une nouvelle fois de d\'etecter la non faisabilit\'e de l'instance (la valeur du plus court chemin sur une ressource d\'epasse le majorant du puits) ou de mettre \`a jour le majorant du probl\`eme par la consid\'eration d'un chemin r\'ealisable pour toutes les ressources.
 \item Enfin, les chemins les moins co\^uteux et les moins consommateurs en chaque ressource allant de tout sommet au puits sont calcul\'es.
\end{itemize}
D'une part, l'ensemble des chemins ainsi g\'en\'er\'es permet d'am\'eliorer le majorant en recombinant les plus courts chemins pour trouver des chemins r\'ealisables de bon co\^ut~: le principe consiste tout simplement \`a consid\'erer, pour chaque arc $(i,j)$ du graphe, les chemins r\'esultant d'une concat\'enation d'un plus court chemin de $s$~\`a~$i$, de l'arc $(i,j)$ et d'un plus court chemin de $j$~\`a~$t$.

D'autre part, cela permet de tester l'accessibilit\'e ou la pertinence (en incluant l'information apport\'ee par le majorant) de chaque sommet et de chaque arc pour les supprimer si possible~: pour un sommet $i$ ({\em resp.}, pour un arc $(i,j)$), il suffit que, pour une m\'etrique donn\'ee, la valeur du plus court chemin de $s$ \`a $i$ additionn\'e \`a la valeur du plus court chemin de $i$ \`a $t$ ({\em resp.}, \`a la valeur de l'arc $(i,j)$ plus la valeur du plus court chemin de $j$ \`a $t$) soit plus grande que le majorant en co\^ut ou que le majorant pr\'esent sur le puits pour cette m\'etrique pour que ce sommet ({\em resp.}, cet arc) soit supprim\'e.
Cette proc\'edure peut donc renvoyer un constat de non r\'ealisabilit\'e, un chemin optimal ou un minorant et un majorant, ce dernier pouvant \^etre associ\'e \`a un chemin. Une description de ce traitement est propos\'ee par l'algorithme~\ref{algo-pretraitement}, o\`u l'on suppose disposer de deux proc\'edures $pcc(s\to,f)$ et  $pcc(\to t,f)$ qui permettent de d\'eterminer respectivement les plus courts chemins de $s$ \`a tous les sommets et les plus courts chemins de tous les sommets \`a $t$, relativement \`a la m\'etrique $f$.\\

Dans le cadre d'un sch\'ema de g\'en\'eration de colonnes, ce traitement permet de d\'etecter une instance non r\'ealisable mais aussi d'\'eliminer des sommets et des arcs n'appartenant pas \`a une solution optimale. De plus, d\`es que la valeur du majorant devient n\'egative, l'\'elimination de sommets et d'arcs non optimaux n'est plus n\'ecessairement opportune si l'objectif est de conserver plusieurs solutions interessantes (de co\^ut n\'egatif). Il suffit alors de rendre stricts les tests sur le majorant dans les blocs d'\'elimination des sommets et des arcs. 

\begin{algorithm}\label{algo-pretraitement}
\caption{PRETRAITEMENT~: Proc\'edure de pr\'etraitement}
\Entree{~\\
- $G = (V,A)$ graphe,
		$c:A\rightarrow \mathbb{R}$ fonction de co\^ut sur les arcs, 
		$b\in \mathbb{N}^R$ vecteur des ressources disponibles au puits~\\
}
\BlankLine
\Sortie{$G'$ sous-graphe partiel de $G$~; $U$ et $L$ majorant et minorant du chemin le moins co\^uteux et v\'erifiant les contraintes}
\BlankLine
$V'\leftarrow V;~A'\leftarrow A;~L\leftarrow 0$\;
$U\leftarrow U_0 = C_{max} \times (|V| -1) +1~ \mathrm{avec}~ C_{max} = \displaystyle{\max_{(i,j) \in A} \{c_{ij}}\}$\;
\Repeter{$chg = false$}{
 $chg \leftarrow false$\;
\BlankLine
 \tcp{Calcul des plus courts chemins}
 $\{P_{s \to i}, i \in V'\} \leftarrow pcc(s \to,c)$\;
 $\{P'_{i \to t}, i \in V'\} \leftarrow pcc(\to t,c)$\;
 $\forall r=1,\ldots,R,\ \{P^r_{s \to i}, i \in V'\} \leftarrow pcc(s \to ,t^r)$\;
 $\forall r=1,\ldots,R,\ \{P'^r_{i \to t}, i \in V'\} \leftarrow pcc(\to t, t^r)$\;
\BlankLine
 \tcp{Consid\'eration de $P_{s \to t}$} 
 \lSi{$\nexists P_{s \to t}$}{
  {L'instance n'est pas r\'ealisable; EXIT}\;
 } 
 \SinonSi{$\forall r=1,\ldots,R,~T^r(P_{s \to t}) \leq b^r_t$}{
  {$P_{s \to t}$ est une solution optimale ; EXIT}\;
 }
 \lSinon{{$L \leftarrow C(P_{s \to t})$}\;}
\BlankLine

 \tcp{Consid\'eration de $P^r_{s \to t}$}
  \lSi{$\exists r,\ T^r(P^r_{s \to t}) > b^r_t$}{
   {L'instance n'est pas r\'ealisable; EXIT}\;
  }
  \lSi{$\exists r,\ \forall q=1,\ldots,R,~T^q(P^r_{s \to t}) \leq b_t^q\ \mathrm{et}\ C(P^r_{s \to t}) < U$}{
  	{$U \leftarrow C(P^r_{s \to t})$}\;
  }
\BlankLine
 \tcp{Recombinaison des chemins}
 \PourTous{$(i,j) \in A'$}{
  \lSi{$
\displaystyle{ \exists P\in \{P_{s \to i}\}\cup  \bigcup_{r=1}^{R} \{P'^r_{s \to i}\},\ \exists P'\in \{P'_{j \to t}\}\cup \bigcup_{r=1}^R\{P'^r_{j \to t}\},\ \forall r=1,\ldots,R,~T^r(P)+t^r_{ij}+T^r(P') \leq b_t^r}$ et $C(P)+c_{ij}+C(P') < U$}{
   $U\leftarrow C(P)+c_{ij}+C(P')$\;
  }
 }
\BlankLine
 \tcp{\'Elimination de sommets}
 \PourTous{$i \in V'\setminus \{s,t\}$}{
  \Si{$\exists r=1,\ldots,R,~T^r(P^r_{s \to i})+T^r(P'^r_{i \to t}) > b_t^r$}{
   Supprimer le sommet $i$ et tous ses arcs incidents$;~chg \leftarrow true$\;
  }
  \Si{$C(P_{s \to i})+C(P'_{i \to t}) \geq U$}{
   Supprimer le sommet $i$ et tous ses arcs incidents$;~chg \leftarrow true$\;
  }
 }
\BlankLine
 \tcp{\'Elimination d'arcs}
 \PourTous{$(i,j) \in A'$}{
  \Si{$\exists r=1,\ldots,R,~T^r(P^r_{s \to i})+t_{ij}^r+T^r(P'^r_{j \to t}) > b_t^r$}{
   Supprimer l'arc $(i,j);~chg \leftarrow true$\;
  }
  \SinonSi{$C(P_{s \to i})+c_{ij}+C(P'_{j \to t}) \geq U$}{
   Supprimer l'arc $(i,j);~chg \leftarrow true$\;
  } 
 }
}
Retourner les bornes $L$ et $U$ ainsi que le chemin correspondant \`a la borne $U$\;

\end{algorithm}

\section{R\'esolution exacte du PCCC}\label{sec-resol} %%  par la programmation dynamique
Nous rappelons que le probl\`eme PCCC est $\mathbf{NP}$-dur, et ce m\^eme pour une seule ressource. Le probl\`eme de d\'ecision associ\'e \`a l'existence d'un chemin de co\^ut inf\'erieur \`a une borne est lui-m\^eme $\mathbf{NP}$-complet pour le cas de deux ressources ou plus. La r\'esolution exacte de ce probl\`eme peut \^etre men\'ee par la programmation dynamique.

\subsection{Cas g\'en\'eral}\label{sub-resol-progdyn}
La programmation dynamique pour ce probl\`eme permet d'\'elaborer un algorithme pseudo-polynomial\footnote{La complexit\'e d\'epend polynomialement de la taille des instances ainsi que des donn\'ees num\'eriques.} de r\'esolution exacte. La programmation dynamique se fonde sur le principe d'optimalit\'e de Bellman~: toute sous-s\'equence d'une s\'equence optimale est optimale. Cela est directement appliquable au plus court chemin~: si $P=\{(s,v_1),\ldots,(v_q,t)\}$ est un plus court chemin de $s$ \`a $t$, alors $P'=\{(s,v_1),\ldots,(v_{i-1},v_i)\}$ est n\'ecessairement un plus court chemin de $s$ \`a $v_i$ pour $i \leq q$.  Avec l'introduction des contraintes de ressource, on ne peut plus adapter directement ce principe pour propager les meilleurs chemins, puisqu'il faut nuancer la notion d'optimalit\'e des sous-s\'equences par la consid\'eration d'un niveau de consommation de ressource~: si $P=\{(s,v_1),\ldots,(v_q,t)\}$ est un plus court chemin de $s$ \`a $t$ consommant $(B^1,\ldots,B^R)$ unit\'es des $R$ ressources, alors  $P'=\{(s,v_1),\ldots,(v_{i-1},v_i)\}$ est un plus court chemin de $s$ \`a $v_i$, parmi les chemins ne consommant pas plus de $(B^1-T^1(P_{i\rightarrow t}),\ldots,B^R-T^R(P_{i\rightarrow t}))$ unit\'es des ressources. Pour g\'erer les niveaux de consommation de co\^ut et de ressource des sous-s\'equences, la programmation dynamique utilise la notion d'\'etiquettes.
\begin{defs}
\'Etiquette\\
Une \'etiquette est un vecteur repr\'esentant un chemin et dont les coordonn\'ees sont le co\^ut et les consommations des diff\'erentes ressources.\\
$E = (E^0, E^1, \ldots ,E^R)$ o\`u $E^0$ est le co\^ut du chemin et $E^r$ pour $r\in \{1,\ldots ,R\}$ est la consommation de la ressource $r$. 
\end{defs}
\begin{rem}
~\\
A tout chemin r\'ealisable entre $s$ et tout sommet correspond une \'etiquette.
\end{rem}
Les diff\'erents algorithmes de programmation dynamique utilisent ces \'etiquettes, mais il n'est pas forc\'ement n\'ecessaire de toutes les garder~: conserver toutes les \'etiquettes reviendrait \`a \'enum\'erer tous les chemins du graphe. Pour \'eliminer les \'etiquettes inutiles, une relation de dominance est d\'efinie.

\begin{defs}
Relation de dominance~:\\
$\begin{array}{ccll}
Une\ \acute{e}tiquette\ E\ domine\ une\ \acute{e}tiquette\ E' &\ ssi\ & & \forall r\in\{0,\ldots,R\},~E^r \leq E'^r\\
(E \succ E') & &et& \exists r\in\{0,\ldots,R\},~E^r < E'^r\\
\end{array}$
\end{defs}
Cette relation de dominance induit un ordre partiel appel\'e ordre de Pareto. Toutes les \'etiquettes n'\'etant pas forc\'ement deux \`a deux comparables, cela permet de d\'efinir un ensemble de majorants pour cet ordre.
\begin{defs}
\'Element Pareto-optimal ou non domin\'e :\\
$E$ est Pareto-optimal $ssi\ \nexists E',\ E' \succ E$
\end{defs}
\begin{defs}
Ensemble Pareto-optimal :\\
Un ensemble Pareto-optimal est un ensemble d'\'el\'ements non domin\'es.\\
$$Pareto =\{E\ |\ \nexists E',\ E' \succ E\}$$
\end{defs}
Cette relation de dominance permet de ne g\'en\'erer que les chemins Pareto-optimaux. Pour $k \in \mathbb{N}$, la conservation de toutes les \'etiquettes non domin\'ees par $k$ autres permet de s'assurer de trouver les $k$ meilleures solutions et permet ainsi, dans notre sch\'ema de g\'en\'eration de colonnes, d'ins\'erer plusieurs colonnes lors d'une m\^eme it\'eration.

Deux types d'algorithmes de programmation dynamique existent~: algorithmes \`a correction d'\'etiquettes (\cite{Desrosiers1983}) et algorithmes \`a fixation d'\'etiquettes (\cite{Desrochers1988}).

\subsubsection{Algorithme \`a correction d'\'etiquettes \cite{Desrosiers1983}}\label{subsub-resol-progdyn-corr}
~\\
Cet algorithme se nomme ainsi car, \`a chaque it\'eration, il va essayer d'am\'eliorer les \'etiquettes d\'ej\`a existantes. En effet, une liste de sommets sur lesquels il existe des \'etiquettes non encore trait\'ees est maintenue. \`A chaque \'etape, un sommet de cette liste est choisi et les \'etiquettes de ce sommet sont propag\'ees aux successeurs de ce sommet. Pour chacune des \'etiquettes ainsi cr\'e\'ees, un test de dominance est effectu\'e pour \'eliminer les \'etiquettes domin\'ees ou d\'ej\`a obtenues. Si une nouvelle \'etiquette appara\^it, le sommet sur lequel cette \'etiquette est pr\'esente est ajout\'e \`a la liste des sommets \`a traiter.

\begin{algorithm}\label{algo-correction}
\caption{CORRECTION~: Algorithme \`a correction d'\'etiquettes}
\Entree{~\\
- $G = (V,A)$~: graphe avec R contraintes de ressource\\
- $Pareto(\mathscr{E})$ renvoie l'ensemble des \'etiquettes non domin\'es de $\mathscr{E}$}
\BlankLine
\Sortie{Chemin contraint de co\^ ut optimal}
\BlankLine

\tcp{$ETIQ(i)$ est l'ensemble des \'etiquettes du sommet $i$, les \'etiquettes \'etant des vecteurs de taille $(1+R)$}

\PourTous{$i \in V$}{$ETIQ(i) \leftarrow \emptyset$\;}
$LIST \leftarrow \{s\}$\;
$ETIQ(s) \leftarrow \{0\}$

\Tq{$LIST \neq \emptyset$}{
Choisir $i \in LIST;~LIST \leftarrow LIST \setminus \{i\}$\;
\PourTous{$j \in succ(i)$}{
	\PourTous{$E \in ETIQ(i)$}{
		\Si{$\forall r \in \{1, \ldots, R\},~E^r+t^r_{ij} \leq b^r_j$}{
			$E'\leftarrow(E^0+c_{ij},E^r+t^r_{ij}, \forall r \in \{1, \ldots, R\})$\;
			$ETIQ(j) \leftarrow Pareto(ETIQ(j) \cup \{E'\})$\;
			\lSi{$E' \in ETIQ(j)$}{$LIST \leftarrow LIST \cup \{j\}$\;}
		}
	}
}
Retourner le chemin ayant le plus petit co\^ut en $t$\; 
} 
\end{algorithm}

\subsubsection{Algorithme \`a fixation d'\'etiquettes \cite{Desrosiers1983}}\label{subsub-resol-progdyn-fix}
~\\
Cet algorithme se nomme ainsi car \`a chaque it\'eration, il va fixer une \'etiquette qui ne pourra plus \^etre modifi\'ee. En effet, une liste d'\'etiquettes non encore trait\'ees est maintenue. \`A chaque \'etape, une \'etiquette (une des non domin\'ees de la liste) est choisie pour \^etre propag\'ee. Les nouvelles \'etiquettes cr\'e\'ees, si elles ne sont pas domin\'ees sur leur sommet, sont rajout\'ees \`a la liste.

\begin{algorithm}\label{algo-fixation}
\caption{FIXATION~: Algorithme \`a fixation d'\'etiquettes}
\Entree{~\\
- $G = (V,A)$~: graphe avec R contraintes de ressource\\
- $Pareto(\mathscr{E})$ renvoie l'ensemble des \'etiquettes non domin\'es de $\mathscr{E}$\\
- $Sommet(E)$ renvoie le sommet d'\'etiquette $E$\\
- $min\_ordre\_lex(\mathscr{E})$ renvoie un \'el\'ement de l'ensemble $\mathscr{E}$ minimum pour l'ordre lexicographique}
\BlankLine
\Sortie{Chemin contraint de co\^ ut optimal}
\BlankLine

\tcp{Les \'etiquettes sont des vecteurs de taille $(1+R)$.}
\tcp{$ETIQ(i)$ est l'ensemble des \'etiquettes du sommet $i$.}
\tcp{$DEF\_ETIQ(i)$ est l'ensemble des \'etiquettes d\'efinitives du sommet $i$.}

\lPourTous{$i \in V$}{$ETIQ(i) \leftarrow \emptyset;~DEF\_ETIQ(i) \leftarrow \emptyset$\;}
\BlankLine
$ETIQ(s) \leftarrow \{(0,\ldots,0)\}$
\BlankLine
\Tq{$\displaystyle{\displaystyle \bigcup_{i \in V} (ETIQ(i) \setminus DEF\_ETIQ(i)) \neq \emptyset}$}{
  $E \leftarrow min\_ordre\_lex(\displaystyle \bigcup_{i \in V} (ETIQ(i) \setminus DEF\_ETIQ(i)))$\;
  $i \leftarrow sommet(E)$\;
  \PourTous{$j \in succ(i)$}{
	  \Si{$\forall \ell \in \{1, \ldots, R\},~E^\ell+t^\ell_{ij} \leq b^\ell_j$}{
			$E'\leftarrow(E^0+c_{ij},E^1+t^1_{ij},\ldots,E^R+t^R_{ij})$\;
			$ETIQ(j) \leftarrow Pareto(ETIQ(j) \cup \{E'\})$\;
	  }
  }
  $DEF\_ETIQ(i) \leftarrow DEF\_ETIQ(i) \cup \{E\}$\;
}
Retourner le chemin ayant le plus petit co\^ut en $t$\; 
\end{algorithm}

\subsection{Cas d'un graphe acyclique}\label{sub-resol-acycl}
~\\
Dans le cas d'un graphe acyclique, un ordre topologique sur les sommets peut \^etre calcul\'e. L'ensemble des sommets peut alors \^etre num\'erot\'e de $1$ \`a $n$ de sorte que le sommet $1$ soit la source, le sommet $n$ le puits et que les sommets de tout arc $(i,j)$ v\'erifie la relation $i<j$. Cet algorithme de programmation dynamique pour les graphes acycliques va appliquer l'algorithme \`a correction d'\'etiquettes vu pr\'ec\'edemment en parcourant les sommets dans l'ordre topologique. Ainsi, lors du traitement d'un sommet, tous les sommets qui le pr\'ec\`edent dans l'ordre topologique ont leurs \'etiquettes d\'efinitives. Le calcul des \'etiquettes de ce sommet ne se fait donc qu'une seule fois.

\begin{algorithm}\label{algo-acyclique}
\caption{ACYCLIQUE~: Algorithme pour les graphes acycliques}
%\begin{algorithmic}
\Entree{~\\
- $G = (V,A)$~: graphe acyclique avec R contraintes de ressource, $V = \{1 \ldots n\}$\\
- $Pareto(\mathscr{E})$ renvoie l'ensemble des \'etiquettes non domin\'es de $\mathscr{E}$}
\BlankLine
\Sortie{Chemin contraint de co\^ ut optimal}
\BlankLine
\tcp{$ETIQ(i)$ est l'ensemble des \'etiquettes du sommet $i$, les \'etiquettes sont des vecteurs de taille $1+R$}
$ETIQ(1) \leftarrow \{(0,\ldots,0)\}$\;
\PourTous{$i$ allant de $2$ \`a $n$}{
	$ETIQ(i) \leftarrow \emptyset$\;
	\PourTous{$j \in pred(i)$}{
		\PourTous{$E \in ETIQ(j)$}{
			\Si{$\forall \ell \in \{1, \ldots, R\},~E^\ell+t^\ell_{ji} \leq b^\ell_j$}{
    			$E'\leftarrow(E^0+c_{ij},E^1+t^1_{ij},\ldots,E^R+t^R_{ij})$\;
				$ETIQ(i) \leftarrow ETIQ(i) \cup \{E'\})$\;
			}
		}
	}
	$ETIQ(i) \leftarrow Pareto(ETIQ(i))$\;		
}
Retourner le chemin ayant le plus petit co\^ut au sommet $n$\;
\end{algorithm}

\subsubsection{Etude de complexit\'e}\label{sub-resol-acycl-complex}
~\\
Pour calculer la complexit\'e en espace, il suffit de calculer le nombre maximum d'\'etiquettes que peut g\'en\'erer cet algorithme. En supposant que chaque sommet est reli\'e \`a tous les sommets le pr\'ec\'edant dans l'ordre topologique, le nombre maximum d'\'etiquettes pour le sommet $i$ ($N_i$) est \'egal au nombre de toutes les \'etiquettes des sommets de plus petit indice ~; la source n'ayant qu'une seule \'etiquette (nulle)~:
$$
\begin{array}{rcl}
 N_1 = 1 & et & \displaystyle N_i = \sum_{j=1}^{i-1} N_j,\ i > 1 \\
\mathrm{soit,}\ N_1 = 1 & et & N_i = 2^{i-2},\ i > 1 
\end{array}
$$
Pour trouver le nombre d'\'etiquettes total, il suffit de faire la somme sur tous les sommets~:
$$
\displaystyle \sum_{i=1}^n N_i = \displaystyle 1 + \sum_{i=2}^n 2^{i-2} =  2^{n-1} = N_{n+1}
$$
Sachant que chaque \'etiquette est un vecteur de taille $1+R$, la complexit\'e en espace est donc en $\mathcal{O}(R\ e^n)$.

La complexit\'e en temps de l'algorithme pr\'ec\'edent se calcule en prenant le cas du sommet 2 \`a part car il n'y a qu'une seule \'etiquette cr\'e\'ee et donc pas de dominance ($i$ it\`ere les sommets, $j$ les pr\'ed\'ecesseurs, $k$ les \'etiquettes, $\ell$ et $m$ sont deux it\'erateurs pour faire les comparaisons deux \`a deux lors de la dominance)~:
$$
\begin{array}{rcl}
  \displaystyle R + \sum_{i=2}^n \left(\sum_{j=1}^{i-1} N_j R + \textrm{C}^2_{N_i} R \right)
  	& = & 
  	\displaystyle R + N_2 R + \textrm{C}^2_{N_2} R + R \sum_{i=3}^n \left(\sum_{j=1}^{i-1} N_j + \textrm{C}^2_{N_i}\right) \\
  \multicolumn{3}{l}{\textrm{en sachant que }N_1 = 1,\ N_2 = 1 \textrm{ et } N_j = 2^{j-2},\ j\leq 2,\ \textrm{on obtient :}}\\
  & = & 
  	\displaystyle 2R + R \sum_{i=3}^n \left(N_i + \frac{N_i\ (N_i-1)}{2}\right) \\
  & = & 
  	\displaystyle 2R + R \sum_{i=3}^n \left(\frac{N_i^2}{2} + \frac{N_i}{2} \right) \\
  & = & 
  	\displaystyle 2R + R \sum_{i=3}^n \left(2^{2i-5} + 2^{i-3} \right) \\
  & = & 
  	\displaystyle R \left(2 + \frac{2\ 4^{n-2}}{3} - \frac{2}{3} + 2^{n-2} - 1 \right)\\
 & = & 
  	\displaystyle R \left(\frac{4^n}{24} + \frac{2^n}{4} + \frac{1}{3} \right)\\
  \displaystyle R + \sum_{i=3}^n \left(\sum_{j=1}^{i-1} \sum_{k=1}^{N_j} R + \sum_{\ell=1}^{N_i-1}\sum_{m=\ell+1}^{N_i} R \right)
  	& = & 
  	\displaystyle \mathcal{O}(R\ e^n)\\
\end{array}
$$

L'algorithme de programmation dynamique propos\'e pour les graphes acycliques n'est pas polynomial. Les algorithmes pour les graphes g\'en\'eraux ne le sont a fortiori pas non plus. Au vu de ces r\'esultats de complexit\'e, l'\'elagage du graphe a priori en supprimant des sommets ou des arcs est primordial pour r\'eduire le nombre de chemins possibles~; de m\^eme, diminuer l'amplitude des fen\^etres de temps et trouver un bon majorant permet aussi de diminuer le nombre d'\'etiquettes calcul\'ees.

\subsection{$k$ Plus Courts Chemins}\label{sub-resol-kpcc}
Des algorithmes polynomiaux exacts ont \'egalement \'et\'e d\'evelopp\'es pour d\'eterminer, non plus {\em un} plus court chemin, mais les $k$ plus courts chemins de $s$ \`a $t$ (ou de $s$ \`a tout autre sommet). Le meilleur algorithme connu \`a ce jour est celui de Eppstein, \cite{Eppstein1998}, de complexit\'e $\mathcal{O}(m + n\, \log(n) + k)$. Il consiste \`a fabriquer un tas des chemins du graphe stock\'es sous forme implicite et tri\'e selon un certain crit\`ere. L'explicitation d'un chemin du tas se fait ensuite en $\mathcal{O}(p)$ o\`u $p$ est la longueur du chemin \`a \'enum\'erer.

Dans le cas du PCCC, cet algorithme peut calculer les consommations de ressource s'additionnant le long de tous les chemins lors de la construction du tas avec une complexit\'e d'ordre $\mathcal{O}(m + Rn\, \log(n))$. La r\'ecup\'eration de la valeur d'une consommations se fait en temps constant. Ainsi, il est possible de tester la r\'ealisabilit\'e d'un chemin en temps $\mathcal{O}(R)$. 

Pour r\'esoudre le probl\`eme par cet algorithme, il faudrait g\'en\'erer le tas de tous les chemins du graphe selon le crit\`ere de co\^ut initial, puis d\'epiler les chemins (dont le nombre total est exponentiel) jusqu'\`a trouver un chemin r\'ealisable, le premier chemin trouv\'e \'etant le chemin optimal. Une telle proc\'edure, de complexit\'e au pire des cas d'ordre exponentiel en temps, pourrait n\'eanmoins consituter une alternative \`a la r\'esolution exacte par programmation dynamique.

%%TODO : VOIR OL Notons cependant que cela n'est possible que dans le cas sp\'ecifique o\`u les contraintes de ressource ne sont pr\'esentes qu'au sommet puits, car les consommations de ressource sont calcul\'ees sur le chemin entier et ne peuvent donc \^etre v\'erifi\'ees en chaque sommet.

\section{R\'esolution approch\'ee du PCCC}\label{sec-approx}

Le probl\`eme du PCCC \'etant $\mathbf{NP}$-difficile, la recherche s'est dirig\'ee vers la r\'esolution approch\'ee du probl\`eme. Quelques heuristiques existent : aggr\'egation de contraintes de ressource \cite{Nagih2005}, recherche de solutions $\varepsilon$-r\'ealisable \cite{Avella2002}. Cette section pr\'esentera un sch\'ema d'approximation pour le cas d'une ressource finale puis une g\'en\'eralisation de ce sch\'ema dans le cas de plusieurs ressources.

\subsection{Sch\'ema d'approximation totalement polynomial}\label{sub-approx-fptas}
Un FPTAS ({\em Fully Polynomial Time Approximation Scheme} ou sch\'ema d'approximation totalement polynomial en temps) a tout d'abord \'et\'e pr\'esent\'e par Hassin \cite{Hassin1992} et ensuite am\'elior\'e par diff\'erents auteurs \cite{Phillips1993,Lorenz2001,Ergun2002}. Ce FPTAS se base sur la programmation dynamique avec la technique {\em d'\'echelonnage et d'arrondis} pour trouver une approximation de la solution. Ce FPTAS est utilisable dans le cas de graphes acycliques, avec une seule contrainte de ressource finale. Un ordre topologique sur les sommets est calcul\'e~; l'ensemble des sommets est donc dor\'enavant num\'erot\'e de $1$ \`a $n$ sachant que le sommet $1$ est la source, le sommet $n$ le puits et tout arc $(i,j)$ v\'erifie la relation $i<j$.

\subsubsection{Sch\'ema initial}\label{subsub-approx-fptas-hassin}
La proc\'edure de programmation dynamique sous-jacente est bas\'ee sur le co\^ut et non sur la ressource, comme c'est le cas habituellement~: on cherche \`a d\'eterminer pour $c=1, 2, \ldots, g_j(c)$ la plus petite consommation de ressource des chemins allant de $1$ \`a $j$ en au plus $c$ unit\'es de co\^ut~; la recherche porte alors sur $g_n(c)$, $c$ \'etant optimal d\`es lors que $g_n(c) \leq b_n$ (i.e., la valeur optimale est le plus petit $c$ pour lequel la consommation de ressource est inf\'erieure \`a la borne du puits). L'utilisation de cette proc\'edure induit donc que les co\^uts sur les arcs sont des entiers strictement positifs.

\begin{algorithm}\label{algo-exact}
\caption{EXACT~: Programmation dynamique bas\'ee sur le co\^ut}
\Entree{Un graphe G acyclique}
\Sortie{Un chemin contraint de co\^ut optimal ($OPT$)}
\BlankLine

{$c\leftarrow 0$}\;

{$g_1(c) \leftarrow 0$}\;

\lPourTous{$j \in \{2,\ldots,n\}$}{$g_j(c)\leftarrow\infty$}\;

\Tq{$g_n(c) > b_n$}
{
	{$c \leftarrow c + 1$}\;
	\PourTous{$j \in \{2,\ldots,n\}$}
	{
	 	$g_j(c)\leftarrow g_j(c-1)$\;
	 	\PourTous{$i|(i,j)\in A$}
		{
	  		\lSi{$c_{ij} \leq c$}{$\displaystyle{g_j(c)\leftarrow \min \left\{g_j(c),g_i(c-c_{ij})+t_{ij}\right\}}$}\;
		}
	}
}
Retourner le chemin de co\^ut $c$ correspondant \`a la consommation de ressource $g_n(c)$
\end{algorithm}

La complexit\'e de l'algorithme EXACT est $\mathcal{O}(m\ OPT)$ avec $n \ll m$. Il s'agit donc d'un algorithme pseudopolynomial, i.e., dont la complexit\'e est polynomiale en la taille de l'instance prise au sens du nombre d'\'el\'ements de la structure \`a coder, mais exponentielle en le logarithme de ses donn\'ees num\'eriques.\\
Une mani\`ere usuelle de se ramener \`a un ordre polynomial de complexit\'e consiste \`a diminuer consid\'erablement l'ordre de grandeur des donn\'ees num\'eriques~; c'est la technique {\emph{ d'\'echelonnage et d'arrondis}. Bien s\^ur, la polynomialit\'e de l'algorithme appliqu\'e \`a l'instance transform\'ee se gagne au prix de l'optimalit\'e~: les solutions obtenues ne sont plus optimales, mais seulement approch\'ees, pour l'instance initiale.\\
Pour le probl\`eme qui nous concerne, on utilise l'algorithme SCALING qui, \'etant donn\'es une borne $B\in \mathbb{N}$ et un rationnel $\delta\in ]0,n]$,  remplace chaque co\^ut $c_{ij}$ par le co\^ut $\displaystyle \left\lfloor\frac{c_{ij}\ (n-1)}{\delta\ B} \right\rfloor$.  

\begin{algorithm}\label{algo-scaling}
\caption{SCALING~: Proc\'edure d'\'echelonnage et d'arrondis}
\Entree{$I=\left(G=(V,A),\ c:A\rightarrow\mathbb{N}\right)$ graphe arc-valu\'e, $B\in \mathbb{N}$, $\delta\in]0,n]$}
\BlankLine
\Sortie{$\tilde{I}(B,\delta)$ graphe arc-valu\'e}
\BlankLine
$\tilde{V} \leftarrow V$\;
$\tilde{A} \leftarrow \emptyset$\;
\BlankLine
\PourTous{$(i,j) \in A$}{
	\Si{$c_{ij}\leq B$}{
		{$\displaystyle{\tilde{c}_{ij}\leftarrow\left\lfloor\frac{c_{ij}(n-1)}{\delta B}\right\rfloor}$}\;
		{$\tilde{A} \leftarrow \tilde{A} \cup (i,j)$}\;
	}
}
\end{algorithm}

La complexit\'e de la proc\'edure SCALING est en $\displaystyle \mathcal{O}\left(m\, \log\left(\frac{n}{\delta}\right)\right)$ (pour chaque arc, une recherche dichotomique dans l'espace $\displaystyle [0,\frac{n}{\delta}]$). De plus, la proc\'edure EXACT appliqu\'ee \`a l'instance $\tilde{I}(B,\delta)$ n\'ecessiterait un temps $\mathcal{O}(m\ \widetilde{OPT})$, o\`u $\widetilde{OPT}$ d\'esigne la valeur optimale sur $\tilde{I}(B,\delta)$. Or, les valeurs de toute solution $P_{s\rightarrow t}$ sur les instances $I$ et $\tilde{I}(B,\delta)$ sont li\'ees par les relations suivantes~:
$$\begin{array}{lclclcl}
			C(P_{s\rightarrow t})	
	&\leq	&\displaystyle \frac{B\delta}{n-1}\tilde{C}(P_{s\rightarrow t})+B\delta	&\hspace*{1cm}&
			\tilde{C}(P_{s\rightarrow t})	
	&\leq	&\displaystyle \frac{n-1}{B\delta}C(P_{s\rightarrow t})
\end{array}$$

En particulier, si les chemins $\tilde{P}_{s\rightarrow t}$ et $P^*_{s\rightarrow t}$ sont respectivement optimaux pour les instances $\tilde{I}(B,\delta)$ et $I$, alors~:
$$\begin{array}{ccccc}
		\widetilde{OPT}
				&\leq 	&\tilde{C}(P^*_{s\rightarrow t})
				&\leq 	&\displaystyle \frac{n-1}{B\delta}OPT\\[7pt]
C(\tilde{P}_{s\rightarrow t})
				&\leq 	&\displaystyle\frac{B\delta}{n-1}\widetilde{OPT}+B\delta
				&\leq 	&OPT+B\delta
\end{array}$$

On d\'eduit de ces relations~: d'une part, si $OPT$ est \`a rapport polynomial $\rho$ de $B$ (i.e., $OPT\leq \rho B$ o\`u $\rho$ est born\'e par un polyn\^ome en $n$), alors le d\'eroulement de l'algorithme EXACT sur $\tilde{I}(B,\delta)$ devient polynomial~; d'autre part, si $B$ est un minorant de $OPT$  (i.e., $B\leq OPT$), la solution renvoy\'ee est $(1+\delta)$-approch\'ee pour le probl\`eme initial. Il revient donc \`a d\'eterminer en temps polynomial une borne $B$ v\'erifiant $B\leq OPT\leq \rho B$. 

La d\'etermination de $B$ s'effectue par recherche dichotomique dans l'intervalle $[LB,UB]$, o\`u $LB$ et $UB$ d\'esignent respectivement un minorant et un majorant de $OPT$ (par exemple, consid\'erer $LB=1$ et $\displaystyle UB=(n-1)\times\max_{(i,j)\in A}\{c_{ij}\}$). Cette recherche est elle-m\^eme fond\'ee sur la proc\'edure TEST qui, pour une instance $I$ et une borne $B$ donn\'ees, renvoie~:
$$TEST(B,\delta) =\left\{
\begin{array}{ll}
OUI & \textrm{si $OPT \geq B$}\\
NON & \textrm{si $OPT < (1+\delta)B$}
\end{array}
\right.
$$ 

\noindent Une mise en \oe{}uvre de cette proc\'edure consiste \`a appliquer la programmation dynamique EXACT sur l'instance $\tilde{I}(B,\delta)$, en limitant le co\^ut maximum \`a tester \`a $\displaystyle \frac{n-1}{\delta}$.

\begin{algorithm}\label{algo-test}
\caption{TEST~: Proc\'edure de test approch\'ee}
\Entree{$I$ instance, $B$ borne sur le co\^ut objectif, $\delta\in]0,n]$ erreur}
\BlankLine
\Sortie{OUI si $OPT \geq B$ ou chemin $P_{s\rightarrow t}$ de valeur $C(P_{s\rightarrow t})<(1+\delta)B$}
\BlankLine
$\tilde{I}(B,\delta)\leftarrow$SCALING($I,B,\delta$)\;
\BlankLine
$\displaystyle \tilde{B}\leftarrow\left\lfloor \frac{n-1}{\delta}\right\rfloor$\;
\BlankLine

\lPourTous{$c$ de $1$ \`a $\tilde{B}$}{$g_s(c) \leftarrow 0$}\;

\lPourTous{$j \in \{2,\ldots,n\}$}{$g_j(0) \leftarrow \infty$}\;

\Pour{$c$ de $1$ \`a $\tilde{B}$}
{
	\PourTous{$j \in \{2,\ldots,n\}$}
	{
	 	$g_j(c)\leftarrow g_j(c-1)$\;
	 	\PourTous{$i/(i,j)\in A$}
		{
	  		\lSi{$\tilde{c}_{ij} \leq c$}
			{$\displaystyle{g_j(c)\leftarrow \min \left\{g_j(c),g_i(c-\tilde{c}_{ij})+t_{ij}\right\}}$}\;
		}
	}
	 \lSi{$g_n(c) \leq b_n$}{Retourner NON et le chemin trouv\'e}\;
}
Retourner OUI
\end{algorithm}

La complexit\'e de la proc\'edure TEST est domin\'ee par l'\'etape de programmation dynamique, qui s'effectue en temps $\displaystyle \mathcal{O}(\ \frac{m\ n}{\delta})$. Lorsque la proc\'edure TEST r\'epond NON, le chemin renvoy\'e, de valeur au plus $\displaystyle \left(\frac{n-1}{\delta}\right)$ dans l'instance $\tilde{I}(B,\delta)$,  est de valeur au plus $B(1+\delta)$ dans le graphe initial. Si, en revanche, elle r\'epond OUI, c'est que tout chemin r\'ealisable est de valeur au moins $\displaystyle \frac{n-1}{\delta}+1$ sur $\tilde{I}(B,\delta)$, et donc, de valeur au moins $B$ sur $I$. Autrement dit, la d\'ecision exacte sur $\tilde{I}$ se transforme en d\'ecision $\delta$-approch\'ee sur $I$.

$$
\begin{array}{ccccc}
	\tilde{I}
	&\left\{\begin{array}{lclcl}	
		\textrm{OUI}	&\Leftrightarrow	&\widetilde{OPT} >  \tilde{B}\\
		\textrm{NON}	&\Leftrightarrow	&\widetilde{OPT}\leq  \tilde{B}
	\end{array}\right.
	&\Rightarrow
	&I
	&\left\{\begin{array}{l}	
		OPT > B\\
		OPT\leq (1+\delta)B
	\end{array}\right.
\end{array}
$$

La recherche dichotomique DICHO consiste alors, partant d'un intervalle initial $[LB,UB]$, \`a appeler it\'erativement la proc\'edure TEST, jusqu'\`a obtenir un encadrement suffisamment fin de la valeur de $OPT$ ($UB < \rho LB$). Si l'on pose $f(LB,UB)=\delta\in\ ]0,n]$ et $g(LB,UB,\delta)=\sqrt{LB\times UB}$, la complexit\'e de la proc\'edure DICHO est en 
$$\mathcal{O}\left(\log\left(\frac{\log\left(UB/LB\right)}{\log(\rho)}\right)\ \left(\frac{m\ n}{\delta} + \log\left(\log(UB/LB)\right)\right)\right)$$
o\`u $\displaystyle \mathcal{O}\left(\log\left(\frac{\log(UB/LB)}{\log(\rho)}\right)\right)$ estime le nombre de tests n\'ecessaires (recherche dichotomique entre $LB$ et $UB$ dans l'espace logarithmique) et $\mathcal{O}(\log\log(UB/LB))$ le temps de calcul d'une valeur approch\'ee (mais suffisante) de $g(LB,UB,\delta)$. Cette valeur approch\'ee est trouv\'ee en cherchant le premier indice $i$ tel que $\displaystyle 2^{2^i} > \frac{UB}{LB}$~; on prend alors $\displaystyle g(LB,UB,\delta)=LB\ 2^{2^{i-2}}$. 

\begin{algorithm}\label{algo-dicho}
\caption{DICHO~: Recherche de la borne $B$}
\Entree{$LB$ et $UB$ bornes inf\'erieure et sup\'erieure de $OPT$, $\rho$ param\`etre
}
\BlankLine
\Sortie{Une borne $LB$ telle que $LB\leq OPT\leq \rho LB$}
\BlankLine
\Tq{$UB > \rho LB$}{
 $\delta \leftarrow f(LB,UB)$\;
 $B \leftarrow g(LB,UB,\delta)$\;
 \lSi{$TEST(B,\delta)$ r\'epond OUI}{$LB \leftarrow B$}\;
 \lSinon{$UB \leftarrow B(1+\delta)$}\;
}
\end{algorithm}

Le sch\'ema d'approximation totalement polynomial peut maintenant \^etre \'enonc\'e. Pour une instance $I$ et une erreur $\varepsilon\in]0,1[$ donn\'ees, le sch\'ema se d\'ecompose en trois \'etapes~:
\begin{itemize}
 \item Tout d'abord, il calcule un minorant $LB$ et un majorant $UB$ de la valeur du probl\`eme.
 \item Ensuite, il raffine \`a l'aide de DICHO l'encadrement de l'optimum jusq'\`a optenir $LB\leq OPT\leq UB \leq \rho LB,\ \rho \in \mathbb{R}$
 \item Enfin, il d\'etermine le chemin optimal dans l'instance $\tilde{I}(LB,\varepsilon)$ gr\^ace \`a EXACT.
\end{itemize}

\begin{algorithm}\label{algo-schema}
\caption{SCHEMA~: Sch\'ema d'approximation totalement polynomial}

\Entree{$\varepsilon$ l'erreur d'approximation, $\rho$ param\`etre 
}
\BlankLine
\Sortie{Un plus court chemin contraint $(1+\varepsilon)$-approch\'e}
\BlankLine
(1)\ D\'eterminer $LB$ et $UB$ bornes inf\'erieure et sup\'erieure de $OPT$\;
\BlankLine
(2)\ $LB\leftarrow$DICHO($LB,UB,\rho$)\;
\BlankLine
(3)\ $P_{s\rightarrow t}\leftarrow$EXACT($\tilde{I}(LB,\varepsilon)$)\;

\end{algorithm}

La complexit\'e des \'etapes $(2)$ et $(3)$ de cet algorithme est donc de 
	$$\mathcal{O}\left(\log\left(\frac{\log(UB/LB)}{\log(\rho)}\right) \left(\frac{m\ n}{\delta}+\log\log\left(\frac{UB}{LB}\right)\right)\right) + 
	\mathcal{O}\left(m\ \left(\rho\ \frac{n}{\varepsilon}\right)\right)$$

Hassin \cite{Hassin1992} utilise les param\`etres constants $\delta=\varepsilon$ et $\rho=2$ pour la proc\'edure de r\'eduction de l'intervalle $[LB,UB]$~; il initialise par ailleurs les bornes $LB$ et $UB$ respectivement \`a $1$ et $(n-1)C_{max}$. La complexit\'e finale du sch\'ema qu'il propose est ainsi d'ordre~:
$$\mathcal{O}\left(\log\log\left(\frac{UB}{LB}\right)\left(\frac{m\ n}{\epsilon} + \log\log\left(\frac{UB}{LB}\right)\right)\right)$$
qui est bien polynomiale en $(n, m, \log(C_{max}), 1/\epsilon)$.

\subsubsection{Am\'eliorations}\label{subsub-approx-fptas-improve}
Lorenz et Raz \cite{Lorenz2001} ont propos\'e un algorithme polynomial pour trouver des bornes inf\'erieure et sup\'erieure dont le rapport est~$n$. En consid\'erant le graphe r\'eduit $G_i = (V, A_i)$ o\`u $A_i$ est l'ensemble des arcs dont le co\^ut est parmi les $i$ plus faibles, il suffit de trouver le $i$ tel que $G_i$ admet un chemin r\'ealisable en consommation de ressource et $G_{i-1}$ n'en admet pas. La borne inf\'erieure est donc le plus grand co\^ut, not\'e $c_{LR^+}$, sur $G_i$  (tout chemin r\'ealisable emprunte au moins un arc de co\^ut $c_{LR^+}$, sinon il existe un chemin r\'ealisable sur $G_{i-1}$) et la borne sup\'erieure est $n\ c_{LR^+}$ (le plus court chemin en consommation de ressource sur $G_i$ est r\'ealisable, de valeur au plus $n\ c_{LR^+}$).

\begin{algorithm}\label{algo-borne}
\caption{BORNE~: Proc\'edure Bornes}
\Entree{Un graphe G}
\Sortie{Deux bornes $LB$ et $UB$ telles que $LB\leq OPT\leq UB\leq n\ LB$}
\BlankLine
$LR_- \leftarrow 0$\;
$LR^+ \leftarrow m$\;
\Tq{$LR_-<LR^+-1$}
{
 	$\displaystyle{\ell \leftarrow \left\lfloor \frac{LR^++LR_-}{2} \right\rfloor}$\;
\BlankLine
	Calculer $P^\ell_{s\rightarrow t}$ plus court chemin de $1$ \`a $n$ en consommation de ressource dans $G_\ell = (V_\ell, A_\ell)$ o\`u $A_\ell = \{(i,j) \in A | c_{ij} \leq c_\ell\}$\;
 	\Si{$C(P^\ell_{1\rightarrow n})\leq b_n$}
	{
		$LR^+ \leftarrow \ell$\;
	}	
 	\Sinon{$LR_- \leftarrow \ell$\;}
}
$LB \leftarrow c_{LR^+}$\;
\BlankLine
$UB \leftarrow C(P^{LR^+}_{s\rightarrow t})$\;
\end{algorithm}

Cette proc\'edure est en $\mathcal{O}(n\, \log^2(n)\ +\ m\, \log(n))$ (en simplifiant avec $\mathcal{O}(\log m) = \mathcal{O}(\log n)$). De plus, les auteurs mettent en \oe{}uvre le sch\'ema avec la valeur $f(LB,UB)= \delta = 1$ et $\rho = 2$. La complexit\'e totale obtenue est de~:
$$\mathcal{O}\left(m\ n\, \log\log(n)\ +\ \frac{m\ n}{\epsilon}\right)$$

Ergun et al., \cite{Ergun2002} exploitent plus finement la qualit\'e de ces bornes initiales pour am\'eliorer encore la proc\'edure de recherche dichotomique, en rendant dynamique la mise \`a jour des param\`etres $\delta$ et $B$ de cette proc\'edure. Ainsi, ils d\'emontrent que si l'on dispose en entr\'ee de DICHO de bornes $LB$ et $UB$ v\'erifiant $\displaystyle \frac{UB}{LB}\leq n$, alors, en posant~:
$$\begin{array}{lcl}
\delta = f(LB,UB) 			&= &\displaystyle\sqrt{\frac{UB}{LB}} -1\\[7pt]
B      = g(LB,UB,\delta) 	&= &\displaystyle\sqrt{\frac{UB\ LB}{1+\delta}}
\end{array}$$
La complexit\'e de DICHO devient $\mathcal{O}(m\ n)$~; par cons\'equent, le sch\'ema dans sa globalit\'e se d\'eroule en temps~:
$$\mathcal{O}\left(\frac{m\ n}{\varepsilon}\right)$$

\subsubsection{\'Etude de la complexit\'e}
Ergun et al. \cite{Ergun2002} effectuent le calcul de complexit\'e de leur sch\'ema en utilisant les valeurs exactes pour $\delta$ et $B$. Hassin \cite{Hassin1992} utilise une m\'ethode dont la complexit\'e est connue pour trouver une valeur approch\'ee des racines carr\'ees. Ici, l'\'etude commence par choisir de bonnes valeurs approch\'ees des param\`etres du sch\'ema suivant cette m\'ethode avant de faire l'\'etude de la complexit\'e. La recherche des valeurs approch\'ees des racines carr\'ees se fait comme suit :
\begin{itemize}
\itemsep = 0.5ex
\item trouver le premier $i$ tel que $\displaystyle a_i = 2^{2^i} > \frac{UB}{LB}$.
\item $\displaystyle 1 + \delta = a_{i-2}$ d'o\`u $\displaystyle\left(\frac{UB}{LB}\right)^\frac{1}{4}<1+\delta\leq\left(\frac{UB}{LB} \right)^\frac{1}{2}$
\item $\displaystyle B = LB\ a_{i-3}$ d'o\`u $\displaystyle {UB}^\frac{1}{8}{LB}^\frac{7}{8}<B\leq{UB}^\frac{1}{4}{LB}^\frac{3}{4}$
\end{itemize}

\noindent D'apr\`es le r\'esultat de la proc\'edure TEST, les bornes sont mises \`a jour :
\begin{itemize}
\item TEST r\'epond OUI :\\
$LB^+ = B$ et $UB^+ = UB$\\[0.1cm]
Cela donne donc $\displaystyle \left(\frac{UB}{LB}\right)^\frac{3}{4} \leq \frac{{UB}^+}{{LB}^+} < \left(\frac{UB}{LB}\right)^\frac{7}{8}$
~\\
\item TEST r\'epond NON :\\
$LB^+ = LB$ et $UB^+ = (1+\delta)\ B$\\[0.1cm]
Cela donne donc $\displaystyle \left(\frac{UB}{LB}\right)^\frac{3}{8} \leq \frac{{UB}^+}{{LB}^+} < \left(\frac{UB}{LB}\right)^\frac{3}{4}$
\end{itemize}
Donc au pire, la plus petite diminution du rapport donne $\displaystyle \frac{{UB}^+}{{LB}^+} < \left(\frac{UB}{LB}\right)^\frac{7}{8}$.
On note $k$ le nombre de passages dans la boucle \textit{tant que} de la proc\'edure DICHO. Avant l'ex\'ecution de la proc\'edure DICHO, le rapport du majorant sur le minorant est born\'e par $n$ et apr\`es cette proc\'edure, il est de $2$ donc la valeur de $k$ est born\'ee comme suit :
$$
\begin{array}{rcl}
 \displaystyle n^{\left(\left(\frac{7}{8}\right)^k\right)} & \leq & 2\\
 \displaystyle \left(\frac{7}{8}\right)^k\, \log(n) & \leq & \log(2)\\
 \displaystyle k\ \log\left(\frac{7}{8}\right)+ \log\log(n) & \leq & \log\log(2)\\
 \displaystyle k & \geq & \displaystyle \frac{\log\log(n)-\log\log(2)}{\displaystyle \log\left(\frac{8}{7}\right)}\\
 \multicolumn{3}{c}{\displaystyle \textrm{Donc}\ k \leq \left\lceil \frac{\log\log(n)-\log\log(2)}{\displaystyle \log\left(\frac{8}{7}\right)} \right\rceil}
\end{array}
$$
En posant respectivement $\delta_i$, $UB_i$ et $LB_i$ les valeurs de $\delta$, du majorant et du minorant de l'\'etape $i$, la complexit\'e de la proc\'edure DICHO s'\'ecrit donc :
$$\displaystyle \sum_{i=1}^r \mathcal{O}\left(\frac{m\ n}{\delta_i}\right) = \mathcal{O}(m\ n)\ \sum_{i=1}^r \mathcal{O}\left(\frac{1}{\delta_i}\right)$$

On a aussi :
$$
\begin{array}{rcl}
	\displaystyle \left(\frac{LB_i}{UB_i}\right)^\frac{1}{2} \leq \frac{1}{UB_i^\frac{1}{2}LB_i^{-\frac{1}{2}}-1} \leq \frac{1}{\delta_i}
		& < & 
		\displaystyle\frac{1}{UB_i^\frac{1}{4}LB_i^{-\frac{1}{4}}-1}\\
\end{array}
$$
Donc :
$$
\begin{array}{rcl}
	\displaystyle \frac{1}{\delta_i} & < &
		\displaystyle\frac{LB_i^\frac{1}{4}}{UB_i^\frac{1}{4}-LB_i^\frac{1}{4}}\\
	& < &	\displaystyle\left(\frac{LB_i}{UB_i}\right)^\frac{1}{4}\ \frac{1}{\displaystyle1-\frac{LB_i^\frac{1}{4}}{UB_i^\frac{1}{4}}}\\
	& < & 
		\displaystyle\left(\frac{LB_i}{UB_i}\right)^\frac{1}{4}\ \frac{1}{1 - \left(\frac{1}{2}\right)^\frac{1}{4}}\ \textrm{car pour }i \leq k\textrm{, on a }\displaystyle \frac{UB_i}{LB_i} > 2 \textrm{ donc }\frac{LB_i}{UB_i} < \frac{1}{2}\\
\\
	\displaystyle \frac{1}{\delta_i}
		& < &
		\displaystyle\left(\frac{LB_i}{UB_i}\right)^\frac{1}{4}\ \left(2 + 2^\frac{1}{4} + 2^\frac{1}{2} + 2^\frac{3}{4}\right)  
\end{array}
$$

$\displaystyle \sum_{i=1}^r \mathcal{O}\left(\frac{1}{\delta_i}\right) < \sum_{i=1}^r \mathcal{O}\left(\left(\frac{LB_i}{UB_i}\right)^\frac{1}{4}\right)$ et on rappelle que $\displaystyle \frac{{UB}_i}{{LB}_i} < \left(\frac{UB_{i-1}}{LB_{i-1}}\right)^\frac{7}{8}$

Finalement, cela donne :
$$
\begin{array}{rcl}
	\displaystyle \sum_{i=1}^r \left(\frac{LB_i}{UB_i}\right)^\frac{1}{4} 
		& < & 
		\displaystyle \sum_{j=0}^{k-1} \left(\frac{LB_k}{UB_k}\right)^{\frac{1}{4}\ \left(\frac{8}{7}\right)^j}\\
	& < &
		\displaystyle \sum_{j=0}^{k-1} 2^{-\frac{1}{4}\ \left(\frac{8}{7}\right)^j}\ \textrm{car }UB_k \leq 2 LB_k\\
	\multicolumn{3}{l}{\textrm{en faisant tendre }k\textrm{ vers }\infty\textrm{, on obtient }}\\
	\displaystyle \sum_{i=1}^r \left(\frac{LB_i}{UB_i}\right)^\frac{1}{4} 
		& < & 
		\displaystyle2^{-\frac{1}{4}}\ \frac{1}{1-2^{-\frac{1}{28}}} < 35\\
	 
\end{array}
$$
La complexit\'e de la proc\'edure DICHO est bien en $\mathcal{O}(m\ n)$, ce qui permet d'arriver \`a une complexit\'e totale pour le sch\'ema de $\displaystyle \mathcal{O}\left(\frac{m\ n}{\epsilon}\right)$.

\subsection{Approche multicrit\`ere}\label{sub-approx-multiobj}
Si le cas d'une ressource se r\'esoud efficacement par la programmation dynamique, en revanche, que peut-on dire du cas de deux ressources ou plus~? Bien s\^ur, puisque d\'ecider m\^eme s'il existe un chemin r\'ealisable est $\mathbf{NP}$-complet, on ne peut garantir de trouver des solutions approch\'ees en temps polynomial. N\'eanmoins, une mani\`ere de g\'en\'eraliser l'approche pr\'ec\'edente consiste \`a se placer dans le cadre multicrit\`ere. il s'agit d'une relaxation du probl\`eme (on rel\^ache les contraintes de consommation de ressource), mais le fait de conserver la nature multicrit\`ere (les consommations de ressource sont vues comme des crit\`eres \`a optimiser) permet peut-\^etre de mieux parcourir l'ensemble des solutions ou, \`a d\'efaut, d'en dessiner les contours. D\`es le d\'ebut des ann\'ees 80, Hansen s'est int\'ress\'e au probl\`eme de plus court chemin bicrit\`ere en en donnant un FPTAS \cite{Hansen1980}. Pour des probl\`emes plus g\'en\'eraux, Papadimitriou et Yannakakis, \cite{Papadimitriou2000}, ont propos\'e une d\'emonstration g\'eom\'etrique de l'existence d'une fronti\`ere de Pareto approch\'ee de taille polynomiale. Les auteurs ont \'egalement donn\'e des th\'eor\` emes d'existence de proc\'edures polynomiales permettant de construire de telles fronti\`eres, qui s'appliquent notamment \`a notre probl\`eme (ces th\'eor\`emes sont d'ailleurs une forme de g\'en\'eralisation des r\'esultats obtenus par Hassin pour le cas d'une ressource, que l'on peut consid\'erer comme probl\`eme bicrit\`ere). Nous d\'efinissons dans un premier temps le probl\`eme de plus court chemin dans un cadre multicrit\`ere afin d'introduire la notion de fronti\`ere de Pareto approch\'ee. Nous proposons ensuite un algorithme permettant de d\'eterminer une telle fronti\`ere, puis concluons en repla\c{c}ant ces r\'esultats dans le cadre sp\'ecifique de la r\'esolution de \textsc{PCCC}. Dans cette section, on se restreint de nouveau aux contraintes de ressource au puits~; de plus, les graphes sont suppos\'es sans circuit et les donn\'ees num\'eriques rationnelles strictement positives.

\subsubsection{Plus court chemin multicrit\`ere}\label{subsub-approx-multiobj-pres}
Dans l'approche multicrit\`ere, les probl\`emes consid\'er\'es sont la g\'en\'eralisation \`a plusieurs objectifs des probl\`emes d'optimisation classique. Pour le probl\`eme de plus court chemin, il s'agit donc de d\'eterminer, dans un graphe orient\'e $G=(V,A)$, un chemin de $s$ \`a $t$ qui optimise, non pas une fonction de co\^ut $C$, mais un ensemble de fonctions $C^1,\ldots,C^R$. Cette version du probl\`eme est not\'ee \textsc{PCC-M}. Les fonctions \`a optimiser sont toutes suppos\'ees \`a valeur dans $\mathbb{Q}_+^*$. De plus, les crit\`eres peuvent \^etre \`a minimiser ou \`a maximiser (par exemple, un crit\`ere souvent pris en consid\'eration pour la construction de chemins est celui de la longueur des chemins, que l'on cherche \`a maximiser). Ainsi, par la suite, on consid\`erera $R$ fonctions $C^1,\ldots,C^R$ \`a optimiser, sans pr\'esupposer de leur sens d'optimisation (maximiser ou minimiser), %ni de leur forme analytique, ni encore 
ni dans un premier temps de leur interpr\'etation (crit\`ere original du probl\`eme ou relaxation d'une contrainte de ressource). Les ensembles des indices des crit\`eres \`a maximiser et \`a minimiser seront respectivement not\'es $\mathscr{R}^{\max}$ et $\mathscr{R}^{\min}$. Pour tout vecteur $w$ de $\mathbb{R}^{R}$, $w_{|\mathscr{R}^{\max}}$ (resp., $w_{|\mathscr{R}^{\min}}$) d\'esigne sa restriction aux indices de $\mathscr{R}^{\max}$ (resp., de $\mathscr{R}^{\min}$).

La multiplicit\'e des fonctions \`a optimiser fait qu'elles n'induisent plus a priori un ordre complet sur l'ensemble des solutions r\'ealisables~: $(1,2)$ et $(2,1)$ sont par exemple incomparables dans $\mathbf{R}^2$. \'Evidemment, il est toujours possible de d\'efinir malgr\'e tout un ordre complet, par exemple, en consid\'erant l'ordre lexicographique, ou encore, en consid\'erant, lorsque cela est possible, une combinaison lin\'eaire des crit\`eres \`a optimiser~; auquel cas on se ram\`ene \`a un probl\`eme monocrit\`ere. N\'eanmoins, si l'on souhaite conserver la nature multicrit\`ere du probl\`eme (et c'est notre cas~: la ressource~1 ne doit pas plus exc\'eder sa borne que ne le doit la ressource~2), alors il faut travailler sur l'ordre partiel induit par la relation de dominance, d\'ej\`a pr\'esent\'ee en section \ref{sub-resol-progdyn}. On ne cherche plus d\`es lors \`a d\'eterminer une solution optimale, mais un ensemble de solutions {\em efficaces}, ou non domin\'ees, appel\'e fronti\`ere de Pareto.

\begin{defs}\label{def-pareto}
Fronti\`ere de Pareto pour \textsc{PCC-M}~:\\
Soit $I=\left(G=(V,A) ; s,t\in V ; C^1,\ldots,C^R: \mathcal{P}\rightarrow\mathbb{Q}\right)$ une instance de \textsc{PCC-M}, o\`u $\mathcal{P}$ d\'esigne l'ensemble des chemins sur $G$. La fronti\`ere de Pareto de $I$, not\'ee $\mathscr{P}(I)$, est l'ensemble des solutions non domin\'ees de $I$. Formellement, $P(I)$ est l'ensemble des chemins $P_{s \to t}$ qui v\'erifient pour tout chemin ${P'}_{s \to t}\in \mathcal{P}$~:
$$\begin{array}{ll}
		Si &\exists r\in 1,\ldots,R,\ t.q.\ 
			\left\{\begin{array}{llcl}
									&\left(r\in \mathscr{R}^{\max}\right) 	&\wedge	&\left(C^r(P'_{s \to t}) > C^r(P_{s \to t})\right)\\
					\textrm{ou}		&\left(r\in \mathscr{R}^{\min}\right)	&\wedge &\left(C^r(P'_{s \to t}) < C^r(P_{s \to t})\right)
			\end{array}\right.\\[20pt]
		Alors  							&\exists \ell\in 1,\ldots,R,\ t.q.\ 
			\left\{\begin{array}{llcl}
									&\left(\ell\in \mathscr{R}^{\max}\right) 	&\wedge	&\left(C^\ell(P_{s \to t}) > C^\ell(P'_{s \to t})\right)\\
						\textrm{ou}	&\left(\ell\in \mathscr{R}^{\min}\right)		&\wedge &\left(C^\ell(P_{s \to t}) < C^\ell(P'_{s \to t})\right)
						\end{array}\right.
\end{array}$$
\end{defs}

\subsubsection{Fronti\`ere de Pareto $\varepsilon$-approch\'ee}\label{subsub-approx-multiobj-epsilon}
Potentiellement (mais pas n\'ecessairement), que l'on se situe dans l'espace des solutions ou dans celui des valeurs, la fronti\`ere de Pareto est de taille exponentielle~: d\'eterminer cette fronti\`ere n'est donc pas envisageable. En outre, pour certains probl\`emes (et c'est notamment le cas du plus court chemin), d\'ecider m\^eme si un point donn\'e est ou non Pareto optimal est d\'ej\`a $\mathbf{NP}$-complet. Aussi, \`a d\'efaut de manipuler l'ensemble des points Pareto-optimaux, il pourrait \^etre int\'eressant d'exhiber un sous-ensemble de taille raisonnable de points qui permettent de repr\'esenter, de fa\c{c}on approch\'ee mais ma\^{\i}tris\'ee, tous les points de l'ensemble. C'est l\`a la vocation de la fronti\`ere de Pareto $\varepsilon$-approch\'ee.  

%% HERE
\begin{defs}\label{def-epsilonpareto}
Fronti\`ere de Pareto $\varepsilon$-approch\'ee pour \textsc{PCC-M}~:\\
Soit $I=\left(G=(V,A) ; s,t\in V ; C^1,\ldots,C^R: \mathcal{P}\rightarrow\mathbb{Q}\right)$ une instance de \textsc{PCC-M}, o\`u $\mathcal{P}$ d\'esigne l'ensemble des chemins sur $G$. Un sous-ensemble $\mathscr{P}_{\varepsilon}(I)$ de solutions r\'ealisables de $I$ est une fronti\`ere de Pareto $\varepsilon$-approch\'ee si~:
$$\begin{array}{l}
\forall P'_{s \to t}\in\mathcal{P},\ \exists P_{s \to t}\in  \mathscr{P}_{\varepsilon}(I)\ t.q.~:\\[7pt]
\forall r=1,\ldots,R,\ 
\left\{\begin{array}{ll}
		C^r(P'_{s \to t}) < (1+\varepsilon) C^r(P_{s \to t})	&\textrm{si }r\in \mathscr{R}^{\max}\\[7pt]%&\textrm{si l'objectif $C^r$ est \`a maximer}\\
		C^r(P'_{s \to t}) > (1-\varepsilon) C^r(P_{s \to t})	&\textrm{si }r\in \mathscr{R}^{\min}%%\textrm{si l'objectif $C^r$ est \`a minimiser}
		\end{array}\right.
\end{array}$$
\end{defs}

Cette d\'efiniton signifie que tout chemin  $P'_{s \to t}$ a un t\'emoin $P_{s \to t}$ dans $\mathscr{P}_{\varepsilon}(I)$ dont le vecteur de peformance r\'ealise {\em au moins} $1/(1+\varepsilon)C^r(P'_{s \to t})$ si $r\in _{|\mathscr{R}^{\max}}$, {\em au plus} $1/(1-\varepsilon)C^r(P'_{s \to t})$ si $r\in _{|\mathscr{R}^{\min}}$.

%Cette d\'efiniton signifie que tout chemin  $P'_{s \to t}$ a un t\'emoin $P_{s \to t}$ dans $\mathscr{P}_{\varepsilon}(I)$ qui se situe dans le c\^{o}ne ouvert $\mathcal{C}_\varepsilon(P'_{s \to t})$ de $\mathbb{R}^{R}$ d\'efini par~:

%$$\mathcal{C}_\varepsilon(P'_{s \to t})=\left\{\left(\frac
%			{\frac{1}{1+\varepsilon} C^r(P'_{s \to t})_{|\mathscr{R}^{\max}}}
%			{\frac{1}{1-\varepsilon} C^r(P'_{s \to t})_{|\mathscr{R}^{\min}}}\right)
%+ \lambda\cdot \left(\frac{\ \ \, \vec{1}_{|\mathscr{R}^{\max}}}{-\vec{1}_{|\mathscr{R}^{\min}}}\right)\ |\ \lambda > 0^R\right\}$$
%o\`u $\vec{1}$ d\'esigne le vecteur uniform\'ement \'egal \`a~1 sur $\mathbb{R}^{R}$ et o\`u pour tout vecteur $w$ de $\mathbb{R}^{R}$, $w_{|\mathscr{R}^%{\max}}$ (resp., $w_{|\mathscr{R}^{\min}}$) d\'esigne sa restriction aux indices de $\mathscr{R}^{\max}$ (resp., de $\mathscr{R}^{\min}$).

Les auteurs dans \cite{Papadimitriou2000} d\'emontrent que tout probl\`eme multicrit\`ere admet une fronti\`ere de Pareto $\varepsilon$-approch\'ee de taille polynomiale en $|I|$ et en $1/\varepsilon$ (mais exponentielle en $R$). Ce r\'esultat pourrait sembler surprenant, la taille d'une fronti\`ere de Pareto \'etant potentiellement exponentielle~; pourtant, la preuve est relativement simple~; nous la pr\'esentons ici dans le cadre de \textsc{PCC-M}. Si $C^r_{min}$ et $C^r_{maj}$ d\'esignent respectivement un minorant et un majorant de $C^r(P_{s \to t})$ (on rappelle que la valeur $C^r(P_{s \to t})$ est suppos\'ee \^{e}tre rationnelle strictement positive) pour tout $r$, on d\'ecoupe tout d'abord l'intervalle $[C^r_{min},C^r_{maj}]$ en une suite $\left([c^r_i,c^r_{i+1}]\right)_i$ de $H^r$ intervalles, o\`u $c^r_{i+1} = (1+\varepsilon)c^r_{i}$ si $r\in \mathscr{R}^{\max}$, $c^r_{i+1} = (1-\varepsilon)c^r_{i}$ si $r\in \mathscr{R}^{\min}$, pour un param\`etre d'erreur $\varepsilon$. Cette proc\'edure de quadrillage de l'espace des valeurs est pr\'ecis\'ement d\'ecrite dans l'algorithme QUADRILLAGE. Si l'on d\'efinit la quantit\'e $MAJ$ comme $MAJ=\max_{r=1}^R\{C^r_{maj}/C^r_{min}\}$, la complexit\'e de cet algorithme est d'ordre $\mathcal{O}\left(\left(\frac{R}{\min\{\varepsilon_M,\varepsilon_m\}}\log(MAJ)\right)\right)$. Si la description propos\'ee diff\'eriencie les param\`etres d'erreur $\varepsilon_M$ et $\varepsilon_m$ pour les crit\`eres \`a maximiser et \`a minimiser, on suppose pour l'instant~: $\varepsilon_M=\varepsilon_m=\varepsilon$.

\begin{algorithm}\label{algo-quadrillage}
\caption{QUADRILLAGE~: quadrillage de l'espace des valeurs}
\Entree{$I$ instance, $\varepsilon_M,\varepsilon_m\in]0,1[$ erreurs pour les crit\`eres \`a maximiser / minimiser}
\BlankLine
\Sortie{$\cup_{r=1}^R\{(c^r_0,\ldots,c^r_{H^r})\}$ discr\'etisation de l'espace $\otimes_{r=1}^R[C^r_{min},C^r_{maj}]$}
\BlankLine
\PourTous{$r \in \mathscr{R}^{\max}$}
{
	$H^r = \lceil \log_{(1+\varepsilon_M)}\left(C^r_{maj}/C^r_{min}\right)\rceil$\;
	\PourTous{$i \in \{0,\ldots, H^r\}$}
	{
		$c^r_{i} = (1+\varepsilon_M)^iC^r_{min}$\;
	}
}
\PourTous{$r \in \mathscr{R}^{\min}$}
{
	$H^r = \lceil \log_{(1-\varepsilon_m)}\left(C^r_{min}/C^r_{maj}\right)\rceil$\;
	\PourTous{$i \in \{0,\ldots, H^r\}$}
	{
		$c^r_{i} = (1-\varepsilon_m)^iC^r_{maj}$\;
	}
}
\end{algorithm}

Une fois le quadrillage effectu\'e, on consid\`ere chacun des hypercubes $[c^1_{i^1},c^1_{i^1+1}]\times\ldots\times[c^R_{i^R},c^R_{i^R+1}]$ pour $(i^1,\ldots,i^R)\in \{0,\ldots,H^1-1\}\times\ldots\times\{0,\ldots,H^R-1\}$. Il suffit alors de choisir, lorsqu'un tel point existe, un point de $\mathscr{P}_\varepsilon(I)$ par hypercube. L'ensemble $\mathscr{P}_\varepsilon(I)$ ainsi construit est bien une fronti\`ere de Pareto $\varepsilon$-approch\'ee, de taille born\'ee par~:
$$\mathcal{O}\left(\left(\frac{1}{\varepsilon}\log(MAJ)\right)^R\right)$$

En effet, soit $P'_{s \to t}$ un chemin de $s$ \`a $t$~; son \'etiquette $\left(C^1(P'_{s \to t}),\ldots,C^R(P'_{s \to t})\right)$ appartient \`a un hypercube $H$ repr\'esent\'e par le point $\left(c^1_{i^1},\ldots,c^R_{i^R}\right)$ dont les coordonn\'ees sont d\'efinies par~: 
$$\left\{\begin{array}{lclclc}
			c^r_{i^r} 	&\leq &C^r(P'_{s \to t}) &< &c^{r}_{i^r+1} 	&\textrm{si } r\in \mathscr{R}^{\max}\\[7pt]
			c^r_{i^r} 	&\geq &C^r(P'_{s \to t}) &> &c^{r}_{i^r+1} 	&\textrm{si } r\in \mathscr{R}^{\min}
	\end{array}\right.$$

L'hypercube $H$ consid\'er\'e consiste alors en le produit cart\'esien des intervalles $[c^r_{i^r},c^{r}_{i^r+1}[$ pour les indices correspondant \`a un crit\`ere \`a maximiser et $]c^r_{i^r+1},c^{r}_{i^{r}}]$ pour les indices correspondant \`a un crit\`ere \`a minimiser. Il existe n\'ecessairement une \'etiquette $E$ correspondant \`a un chemin de $\mathscr{P}_\varepsilon(I)$ dans $H$ (cette \'etiquette pouvant \^{e}tre celle de $P'_{s \to t}$)~; alors, par construction~:

$$\begin{array}{ll}
c^r_{i^r}\leq E^r\textrm{ et }C^r(P'_{s \to t}) < (1+\varepsilon)c^r_{i^r}\  
	\Rightarrow\ C^r(P'_{s \to t}) < (1+\varepsilon)E^r		&\textrm{si }r\in \mathscr{R}^{\max}\\[10pt]
c^r_{i^r}\geq E^r\textrm{ et }C^r(P'_{s \to t}) > (1-\varepsilon)c^r_{i^r}\  
	\Rightarrow\ C^r(P'_{s \to t}) > (1-\varepsilon)E^r 
&\textrm{si }r\in \mathscr{R}^{\min}
\end{array}$$

La proc\'edure de recherche d'une fronti\`ere de Pareto $\varepsilon$-approch\'ee que nous venons de discuter est formellement d\'ecrite dans l'algorithme PARETO-EPSILON. Cet algorithme n'est cependant pas polynomial, puisqu'on ne sait a priori pas d\'ecider de l'existence d'un chemin dans un hypercube donn\'e. Notons pour finir que cette construction reste valide si l'on choisit pour chaque point $(c^1_{i^1},\ldots,c^R_{i^R})$ un chemin $P'_{s \to t}$ dont l'\'etiquette se situe dans l'hypercube plus large~:
$$\displaystyle{ \otimes_{r\in \mathscr{R}^{\max}}[c^r_{i^r},c^r_{H^r}] \otimes_{r\in \mathscr{R}^{\min}}[c^r_{H^r},c^r_{i^r}]}$$

\begin{algorithm}\label{algo-pareto-epsilon}
\caption{PARETO-EPSILON~: Fronti\`ere de Pareto $\varepsilon$-approch\'ee}
\Entree{$I$ instance, $\varepsilon\in]0,1[$ erreur}
\BlankLine
\Sortie{$\mathscr{P}_\varepsilon(I)$ Fronti\`ere de Pareto $\varepsilon$-approch\'ee}
\BlankLine
$\mathscr{P}_\varepsilon(I) \leftarrow \emptyset$\;
$\cup_{r=1}^R\{(c^r_0,\ldots,c^r_{H^r})\} \leftarrow$ QUADRILLAGE($\varepsilon,\varepsilon$)\;
\PourTous{$(i^1,\ldots,i^R) \in \{0,\ldots,H^1-1\}\times\ldots\times\{0,\ldots,H^R-1\}$}
{
	\Si{$\exists$ un chemin $P_{s \to t}$ \`a valeur dans 
$\otimes_{r\in \mathscr{R}^{\max}}[c^r_{i^r},c^r_{i^r+1}]\otimes_{r\in \mathscr{R}^{\min}}[c^r_{i^r+1},c^r_{i^r}]$}
		{$\mathscr{P}_\varepsilon(I) \leftarrow \mathscr{P}_\varepsilon(I)\cup\{P_{s \to t}\}$\;} 
}
\end{algorithm}

\subsubsection{Sch\'ema complet d'approximation de la fronti\`ere de Pareto}\label{subsub-approx-multiobj-pcc}
Dans \cite{Papadimitriou2000} sont \'enonc\'es des th\'eor\`emes de caract\'erisation de la constructibilit\'e en temps polynomial de fronti\`eres de Pareto $\varepsilon$-approch\'ees (o\`u, par polynomial, on entend polynomial en $|I|$ et $1/\varepsilon$, exponentiel en $R$)~; ces constructions sont qualifi\'ees de {\em sch\'emas complets d'approximation} pour le probl\`eme multicrit\`ere. La version multicrit\`ere d'un probl\`eme d'optimisation admet un sch\'ema complet d'approximation de sa fronti\`ere de Pareto {\em ssi} l'on peut d\'ecider en temps polynomial la version d\'ecision approch\'ee (de type GAP) du probl\`eme multicrit\`ere. En particulier, si les fonctions \`a optimiser sont lin\'eaires, discr\`etes et que la version monocrit\`ere exacte est d\'ecidable en temps pseudopolynomial, alors le probl\`eme multicrit\`ere admet un sch\'ema complet d'approximation. S'agissant du probl\`eme de plus court chemin, on sait par la programmation dynamique d\'ecider en temps pseudopolynomial, pour une instance $I$ et une valeur $B$, s'il existe sur $I$ un chemin de valeur $B$~; on en d\'eduit donc que PCC-M admet un sch\'ema complet d'approximation de sa fronti\`ere de Pareto. Nous proposons ici un tel sch\'ema, qui est une forme de g\'en\'eralisation au cadre multicrit\`ere des algorithmes propos\'es dans la section pr\'ec\'edente.

\paragraph{Principe.}
Dans le cadre monocrit\`ere \`a une ressource, pour une instance $I$ et une borne $B$ donn\'ees, on ne sait pas d\'ecider en temps polynomial s'il existe un chemin de co\^ut au plus $B$ tout en consommant au plus $b_n$ quantit\'e de ressource. En revanche, on peut pour une erreur $\varepsilon$ donn\'ee d\'ecider en temps polynomial (en $I$ et en $1/\varepsilon$) s'il existe un chemin de co\^ut au plus $(1+\varepsilon)B$ et de consommation au plus $b_n$, ou si tout chemin, soit co\^ute au moins $B$, soit consomme strictement plus que $b_n$. Cette notion de test approch\'e se g\'en\'eralise naturellement \`a plusieurs fonctions~: on ne sait pas, pour un point $(c^1_{i^1},\ldots,c^R_{i^R})$ donn\'e, d\'ecider s'il existe un chemin $P_{s \to t}$ dont l'\'etiquette $(C^1(P_{s \to t}),\ldots,C^R(P_{s \to t}))$ v\'erifie pour tout crit\`ere $C^r(P_{s \to t})\geq c^r_{i^r}$ si $r\in \mathscr{R}^{\max}$, $C^r(P_{s \to t})\leq c^r_{i^r}$ si $r\in \mathscr{R}^{\min}$, ou si tout chemin n'atteint pas ou exc\`ede strictement l'une au moins des bornes $c^r_{i^r}$. En revanche, on peut d\'ecider en temps polynomial pour des erreurs $\varepsilon_M$ et $\varepsilon_m$ donn\'ees s'il existe un chemin respectant ces bornes, ou si tout chemin est en-de\c{c}a de $(1+\varepsilon_M)c^r_{i^r}$ pour un certain crit\`ere \`a maximiser ou au-del\`a de  $(1-\varepsilon_m)c^r_{i^r}$ pour un certain crit\`ere \`a minimiser. De nouveau, ce test approch\'e est obtenu en travaillant sur une instance $\tilde{I}$ modifi\'ee~: on ram\`ene les donn\'ees num\'eriques de l'instance $I$ initiale \`a un ordre polynomial par une proc\'edure d'\'echelonnage et d'arrondis d\'ependant du point $(c^1_{i^1},\ldots,c^r_{i^r})$ et des degr\'es de pr\'ecision $\varepsilon_M,\varepsilon_m$~; au prix de l'erreur introduite, on gagne le d\'eroulement polynomial sur $\tilde{I}$ des algorithmes exacts qui sont de complexit\'e th\'eorique pseudopolynomiale.

\paragraph{Algorithme de d\'ecision exacte.}
Nous pr\'esentons ici l'algorithme ACYCLIQUE-M qui est une adaptation au cadre multicrit\`ere de l'algorithme \`a fixation d'\'etiquette \ref{algo-acyclique}. Cet algorithme renvoie, pour une instance $I$ de PCC-M et un vecteur $B$ de $\mathbb{Q}^R$, l'ensemble $\mathscr{E}$ des \'etiquettes $E\in \mathbb{Q}^R$ non domin\'ees v\'erifiant~:
$$E_{|\mathscr{R}^{\max}} \geq B_{|\mathscr{R}^{\max}}\textrm{ et }E_{|\mathscr{R}^{\min}} \leq B_{|\mathscr{R}^{\min}}$$ 

ACYCLIQUE-M  renvoie \'egalement les chemins associ\'es aux \'etiquettes de $\mathscr{E}$, en ne conservant toutefois qu'un chemin par \'etiquette. Selon l'analyse faite en section \ref{sub-resol-acycl-complex}, la complexit\'e de cet algorithme est d'ordre~: 
$$R + \sum_{i=2}^n \left(\sum_{j=1}^{i-1} N_j R + \textrm{C}^2_{N_i} R \right) + N_n R$$

Or, si les donn\'ees sont enti\`eres, si $M$ d\'esigne un majorant du nombre de valeurs possibles 
(consid\'erer par exemple, si l'on note $C_{maj}= \max_{r=1}^R \{C^r_{maj}\}$ et $C_{min}=\min_{r=1}^R \{C^r_{min}\}$, $M=(n-1)(C_{maj}-C_{min}+1)\leq \mathcal{O}(nC_{maj})$), le nombre $N_j$ d'\'etiquettes en tout sommet $j$ est born\'e par $M^R$. Si l'on ne conserve qu'un chemin par \'etiquette en chaque sommet, la complexit\'e de ACYCLIQUE-M est alors d'ordre au plus~:
$$\mathcal{O}\left(R\,M^R\left(n^2+nM^R\right)\right) = 
\mathcal{O}\left( R\, n^{2R+1}\, C_{maj}^{2R} \right)$$

L'algorithme ACYCLIQUE-M permet ainsi de d\'ecider en temps pseudopolynomial, pour une instance $I$ de PPC-M aux donn\'ees num\'eriques enti\`eres et un vecteur $B$, s'il existe un chemin dans l'hyperespace  
$\otimes_{r\in \mathscr{R}^{\max}}[B^r,+\infty[\otimes_{r\in \mathscr{R}^{\min}}[0,B^r]$, ou si tout chemin exc\`ede $B^r$ pour un certain crit\`ere $r\in \mathscr{R}^{\min}$, ou n'atteint pas $B^r$ pour un certain crit\`ere $r\in\mathscr{R}^{\max}$.

\begin{algorithm}\label{algo-acyclique-m}
\caption{ACYCLIQUE-M~: Algorithme pour les graphes acycliques}
%\begin{algorithmic}
\Entree{$I=\left(G=(V,A);\ c^1,\ldots,c^R:A\rightarrow\mathbb{Q}\right)$ instance de PPC-M, $B=(B^1,\ldots,B^R)$ vecteur borne}
\BlankLine
\Sortie{Ensemble des \'etiquettes $(E^1,\ldots,E^R)$ non domin\'ees v\'erifiant 
	$\left\{\begin{array}{ll}
		E^r\geq B^r	&\forall r\in \mathscr{R}^{\max}\\
		E^r\leq B^r	&\forall r\in \mathscr{R}^{\min}
		\end{array}\right.$}
\BlankLine
\tcp{$ETIQ(i)$ est l'ensemble des \'etiquettes du sommet $i$}
$ETIQ(1) \leftarrow (0)$
\BlankLine
\PourTous{$i$ allant de $2$ \`a $n-1$}{
	$ETIQ(i) \leftarrow \emptyset$\;
	\PourTous{$j \in predecesseur(i)$}{
		\PourTous{$E \in ETIQ(j)$}{
			\Si{$\forall r \in \mathscr{R}^{\min},~E^r+c^r_{ji} \leq B^r$}{
				$E'\leftarrow(E^r+c^r_{ji}, \forall r \in \{1, \ldots, R\})$\;
				$ETIQ(i) \leftarrow ETIQ(i) \cup \{E'\}$\;
			}
		}
	}
	$ETIQ(i) \leftarrow Pareto(ETIQ(i))$\;
}

\BlankLine
	$ETIQ(n) \leftarrow \emptyset$\;
	\PourTous{$j \in predecesseur(n)$}{
		\PourTous{$E \in ETIQ(j)$}{
			\Si{$\forall r \in \mathscr{R}^{\min},~E^r+c^r_{jn} \leq B^r$ et $\forall r \in \mathscr{R}^{\max},~E^r+c^r_{jn} \geq B^r$}{
				$E'\leftarrow(E^r+c^r_{jn}, \forall r \in \{1, \ldots, R\})$\;
				$ETIQ(n) \leftarrow ETIQ(n) \cup \{E'\}$\;
			}
		}
	}
	$ETIQ(n) \leftarrow Pareto(ETIQ(n))$\;
\BlankLine
Retourner $ETIQ(n)$ \;
\end{algorithm}

\paragraph{Proc\'edures d'\'echelonnage et d'arrondis et de test approch\'ee.} 
\begin{algorithm}\label{algo-scaling-m}
\caption{SCALING-M~: Proc\'edure d'\'echelonnage et d'arrondis multicrit\`ere}
\Entree{$I=\left(G=(V,A);\ c^1,\ldots,c^R:A\rightarrow\mathbb{Q}\right)$ instance de PPC-M, $B\in (\mathbb{Q}_+^*)^R$, $\varepsilon_M,\varepsilon_m\in]0,1[$ erreurs}
\BlankLine
\Sortie{$\tilde{I}(B,\varepsilon_M,\varepsilon_m)$ instance de PPC-M}
\BlankLine
\PourTous{les $(i,j)\in A$}
{
	$\tilde{B}_M \leftarrow \lceil\frac{n}{\varepsilon_M}\rceil$\;
	\PourTous{$r \in \mathscr{R}^{\max}$}
	{
		{$\displaystyle{\tilde{c}^r_{ij}\leftarrow\min
			\left\{
				\left\lfloor\frac{c^r_{ij}n}{\varepsilon_M B^r}\right\rfloor, 	\tilde{B}_M
			\right\}}$}\;}

	$\tilde{B}_m \leftarrow \lfloor\frac{n}{\varepsilon_m}\rfloor$\;
	\PourTous{$r \in \mathscr{R}^{\min}$}
	{
		{$\displaystyle{\tilde{c}^r_{ij}\leftarrow\min
			\left\{
				\left\lceil\frac{c^r_{ij}n}{\varepsilon_m B^r}\right\rceil, \tilde{B}_m
			\right\}}$}\;}
}
\end{algorithm}

Pour se ramener d'un ordre pseudopolynomial \`a un ordre polynomial, on \'echelonne les donn\'ees num\'eriques de l'instance $I$ par l'algorithme SCALING-M dont la complexit\'e est d'ordre :
$$\mathcal{O}\left( mR\, \log\left(\frac{n}{\varepsilon}\right) \right) = \mathcal{O}\left( n^2R\, \log\left(\frac{n}{\varepsilon}\right) \right)$$
L'arrondi effectu\'e pour obtenir des donn\'ees enti\`eres sur  $\tilde{I}$ a pour cons\'equence que la d\'ecision n'est plus exacte, mais seulement approch\'ee, sur l'instance initiale (algorithme TEST-M). Consid\'erons l'instance $\tilde{I}$ obtenue \`a partir de $I$ apr\`es l'appel \`a la proc\'edure SCALING-M pour les param\`etres $B=(c^1_{i^1},\ldots,c^R_{i^R})$, $\varepsilon_M$ et $\varepsilon_m$. On voit facilement que l'on a les relations suivantes entre les \'etiquettes sur les instances $\tilde{I}$ et $I$ d'un chemin $P_{s \to t}$~: 
$$\begin{array}{lcl}
\left\{\begin{array}{ll}
\tilde{C}^r(P_{s \to t})\geq \tilde{B}_M	&\forall r\in \mathscr{R}^{\max}\\[7pt]
\tilde{C}^r(P_{s \to t})\leq \tilde{B}_m	&\forall r\in \mathscr{R}^{\min}
\end{array}\right.	
	&\Rightarrow
		&\left\{\begin{array}{ll}
			C(P_{s \to t})_{|\mathscr{R}^{\max}}\geq B_{|\mathscr{R}^{\max}}\\[7pt]
			C(P_{s \to t})_{|\mathscr{R}^{\min}}\leq B_{|\mathscr{R}^{\max}}
			\end{array}\right.\\[20pt]
\left\{\begin{array}{ll}
\exists r\in \mathscr{R}^{\max}, &\tilde{C}^r(P_{s \to t})< \tilde{B}_M\\[7pt]
\exists r\in \mathscr{R}^{\min}, &\tilde{C}^r(P_{s \to t})> \tilde{B}_m
\end{array}\right.	
	&\Rightarrow
		&\left\{\begin{array}{ll}
			\exists r\in \mathscr{R}^{\max}, &C^r(P_{s \to t})< (1+\varepsilon_M)B^r\\[7pt]
			\exists r\in \mathscr{R}^{\min}, &C^r(P_{s \to t})> (1-\varepsilon_m)B^r
\end{array}\right.\end{array}$$

Par ailleurs, puisque les valeurs num\'eriques sur $\tilde{I}$ sont born\'ees par $\max\{\tilde{B}_M, \tilde{B}_m\}= \mathcal{O}(n/\min\{\varepsilon_M,\varepsilon_m\})$, le d\'eroulement de la proc\'edure ACYCLIQUE-M sur $\tilde{I}$ pour le param\`etre $B$ d\'efini par $B_{|\mathscr{R}^{\max}} \equiv \tilde{B}_M$ et $B_{|\mathscr{R}^{\min}} \equiv \tilde{B}_m$ devient polynomial en $n$ et $1/\varepsilon_M,1/\varepsilon_m$, d'ordre~: 
$$\mathcal{O}\left(\frac{R\, n^{4R+1}} {\min\{\varepsilon_M,\varepsilon_m\}^{2R}}\right)$$

En cons\'equence de ces deux derni\`eres observations, on peut d\'ecider, pour une instance $I$ d'ordre $n$, un point $(c^1_{i^1},\ldots,c^R_{i^R})$ et deux param\`etres d'erreur $\varepsilon_M,\varepsilon_m$, en temps polynomial en $n$ et en $\max\{1/\varepsilon_M,1/\varepsilon_m\}$~:
$$\left\{\begin{array}{ll}
Si			&\exists P_{s \to t}\textrm{ d'\'etiquette dans l'hyperespace }\otimes_{r\in \mathscr{R}^{\max}}[c^r_{i^r},+\infty[\otimes_{r\in \mathscr{R}^{\min}}[0,c^r_{i^r}]\\
\textrm{Ou si}	&\textrm{l'hyperespace }\otimes_{r\in \mathscr{R}^{\max}}[(1+\varepsilon_M)c^r_{i^r},+\infty[\otimes_{r\in \mathscr{R}^{\min}}[0,(1-\varepsilon_m),c^r_{i^r}]\textrm{ est vide.}\end{array}\right.$$

\begin{algorithm}\label{algo-test-m}
\caption{TEST-M~: Proc\'edure de test approch\'ee sur \textsc{PCC-M}}
\Entree{$I$ instance, $B=(B^1,\ldots,B^R)$ vecteur, $\varepsilon_M,\varepsilon_m\in]0,1[$ erreurs}
\BlankLine
\Sortie{\\$\bullet$~OUI et un chemin $P_{s \to t}$ v\'erifiant
	$\left\{\begin{array}{ll}
		C^r(P_{s \to t})\geq B^r	&\forall r\in \mathscr{R}^{\max}\\
		C^r(P_{s \to t})\leq B^r	&\forall r\in \mathscr{R}^{\min}
		\end{array}\right.$
\\$\bullet$~NON si tout chemin $P_{s \to t}$ v\'erifie $C^r(P_{s \to t}) < (1+\varepsilon_M) B^r$ pour un certain crit\`ere $r$ \`a maximiser ou $C^r(P_{s \to t}) > (1-\varepsilon_m) B^r$ pour un certain crit\`ere $r$ \`a minimiser.}
\BlankLine
$\tilde{I}(B,\varepsilon_M,\varepsilon_m)\leftarrow$SCALING-M($I,B,\varepsilon_M,\varepsilon_m$)\;
\BlankLine
$\displaystyle \tilde{B}_M\leftarrow\left\lceil \frac{n}{\varepsilon_M}\right\rceil$; 
$\displaystyle \tilde{B}_m\leftarrow\left\lfloor \frac{n}{\varepsilon_m}\right\rfloor$\;
\BlankLine
$B'\leftarrow(B^r = B_M,\ \forall r\in \mathscr{R}^{\max};\ B^r = B_m,\ \forall r\in \mathscr{R}^{\min})$\;
$ETIQ\leftarrow$ACYCLIQUE-M($\tilde{I},B'$)\;
\Si{$ETIQ\neq \emptyset$}
	{Retourner OUI et un chemin d'\'etiquette dans $ETIQ$}
\Sinon
	{Retourner NON} 
\end{algorithm}

\paragraph{Sch\'ema complet d'approximation}
Le sch\'ema con\c{c}u consiste \`a op\'erer un quadrillage de l'espace des valeurs, puis \`a appeler la proc\'edure de test approch\'e en le  coin de chaque hypercube issu de ce quadrillage. Pr\'ecis\'ement, l'algorithme PARETO-PCC-M sollicite la proc\'edure QUADRILLAGE  pour les erreurs $\varepsilon_M=\sqrt{1+\varepsilon}-1$ et $\varepsilon_m= 1-\sqrt{1-\varepsilon}$ (on ne fait plus le quadrillage sur $\varepsilon$, puisque le test en chaque hypercube n'est pas exact, mais approch\'e). Il appelle ensuite la proc\'edure TEST-M sur chaque coin $(c^1_{i^1},\ldots,c^R_{i^R})$, en invoquant les m\^emes param\`etres d'erreur $\varepsilon_M$ et $\varepsilon_m$. Les chemins trouv\'es par les appels successifs \`a TEST-M constituent la fronti\`ere de Pareto $\varepsilon$-approch\'ee.

\begin{algorithm}\label{algo-pareto-pcc-m}
\caption{PARETO-PCC-M~: Fronti\`ere de Pareto $\varepsilon$-approch\'ee en temps polynomial}
\Entree{$I$ instance, $\varepsilon\in]0,1[$ erreur}
\BlankLine
\Sortie{$\mathscr{P}_\varepsilon(I)$ Fronti\`ere de Pareto $\varepsilon$-approch\'ee}
\BlankLine
$\mathscr{P}_\varepsilon(I) \leftarrow \emptyset$\;
$\varepsilon_M \leftarrow \sqrt{1+\varepsilon}-1$\;
$\varepsilon_m \leftarrow 1-\sqrt{1-\varepsilon}$\;
$\cup_{r=1}^R\{(c^r_0,\ldots,c^r_{H^r})\} \leftarrow$ QUADRILLAGE($I,\ \varepsilon_M,\ \varepsilon_m$)\;
\PourTous{$(i^1,\ldots,i^R) \in \{0,\ldots,H^1-1\}\times\ldots\times\{0,\ldots,H^R-1\}$}
{
	$B\leftarrow (c^1_{i^1},\ldots,c^R_{i^R})$\;
	\Si{TEST-M($I,B,\varepsilon_M,\varepsilon_m$) renvoie un chemin $P_{s \to t}$}
		{$\mathscr{P}_\varepsilon(I) \leftarrow \mathscr{P}_\varepsilon(I)\cup\{P_{s \to t}\}$\;} 
}
\end{algorithm}

Soit $P'_{s \to t}$ un chemin de $s$ \`a $t$ et soit $\left(C^1(P'_{s \to t}),\ldots,C^R(P'_{s \to t})\right)$ son \'etiquette, on consid\`ere le point $B=\left(c^1_{i^1},\ldots,c^R_{i^R}\right)$ caract\'eris\'e par~: 
$$\left\{\begin{array}{lclclc}
			c^r_{i^r} 	&\leq &C^r(P'_{s \to t}) &< &(1+\varepsilon_M)c^{r}_{i^r} 	&\textrm{si } r\in \mathscr{R}^{\max}\\[7pt]
			c^r_{i^r} 	&\geq &C^r(P'_{s \to t}) &> &(1-\varepsilon_m)c^{r}_{i^r} 	&\textrm{si } r\in \mathscr{R}^{\min}
	\end{array}\right.$$

Si TEST-M($I,B,\varepsilon_M,\varepsilon_m$) a renvoy\'e un chemin $P_{s \to t}$, alors ce chemin v\'erifie~:
$$\left\{\begin{array}{ll}
C^r(P_{s \to t})\geq c^r_{i^r}\ \Rightarrow\ C^r(P'_{s \to t})< (1+\varepsilon_M)c^r_{i^r}\leq (1+\varepsilon)C^r(P_{s \to t}) 
&\textrm{si }r\in \mathscr{R}^{\max}\\[7pt]
C^r(P_{s \to t})\leq c^r_{i^r}\ \Rightarrow\ C^r(P'_{s \to t})> (1-\varepsilon_m)c^r_{i^r}\geq (1-\varepsilon)C^r(P_{s \to t}) 
&\textrm{si }r\in \mathscr{R}^{\min}
\end{array}\right.$$

Sinon, l'existence de $P'_{s \to t}$ assure que TEST-M a n\'ecessairement renvoy\'e un chemin $P_{s \to t}$ pour le point $B_{-}$ d\'efini par~:
$$B_{-}^r= c^r_{i^r-1}\ \forall r=1,\ldots,R$$

Ce chemin v\'erifie bien de nouveau~:
$$\left\{\begin{array}{ll}
C^r(P_{s \to t})\geq \frac{1}{1+\varepsilon_M}c^r_{i^r}\ \Rightarrow\ C^r(P'_{s \to t})< (1+\varepsilon_M)^2C^r(P_{s \to t})= (1+\varepsilon)C^r(P_{s \to t}) &\textrm{si }r\in \mathscr{R}^{\max}\\[7pt]
C^r(P_{s \to t})\leq \frac{1}{1-\varepsilon_m}c^r_{i^r}\ \Rightarrow\ C^r(P'_{s \to t})> (1-\varepsilon_m)^2C^r(P_{s \to t})= (1-\varepsilon)C^r(P_{s \to t}) &\textrm{si }r\in \mathscr{R}^{\min}
\end{array}\right.$$

La complexit\'e de l'algorithme PARETO-PCC-M est domin\'ee par celle des appels successifs \`a TEST-M. Le nombre de ces appels est de l'ordre de $\mathcal{O}\left(\left((1/\varepsilon)\log(MAJ)\right)^R\right)$ (consid\'erer que $\sqrt{1+\varepsilon}-1\sim 1/2\varepsilon$, $1-\sqrt{1-\varepsilon}\sim 1/2\varepsilon$). Or, la complexit\'e de TEST-M somme la complexit\'e temporelle de SCALING-M appliqu\'ee \`a $I$ et celle de ACYCLIQUE-M appliqu\'ee \` a $\tilde{I}$.
On obtient au total une complexit\'e d'ordre~:

$$\mathcal{O}\left( \left(\frac{\log(MAJ)}{\varepsilon}\right)^R 
			\left(Rm\, \log\left(\frac{n}{\varepsilon}\right)\, +\, \frac{Rn^{4R+1}}{\varepsilon^{2R}}\right)\right)
			=\mathcal{O}\left( \left(\frac{\log(MAJ)}{\varepsilon}\right)^R \frac{Rn^{4R+1}}{\varepsilon^{2R}}\right)	$$
	
\subsubsection{Fronti\`ere de Pareto et PCCC}\label{subsub-approx-multiobj-p3c}
La vision multicrit\`ere du probl\`eme de plus court chemin contraint \`a $R$ ressources peut s'\'enoncer comme le probl\`eme consistant \`a trouver un chemin $P_{s \to t}$ qui minimise chaque composante du vecteur $\left(C(P_{s \to t}),T^1(P_{s \to t}),\ldots,T^{R}(P_{s \to t})\right)$. Si l'on dispose d'une fronti\`ere de Pareto $\varepsilon$-approch\'ee $\mathscr{P}_\varepsilon$, cela signifie que tout chemin $P'_{s \to t}$ de $s$ \`a $t$ est repr\'esent\'e par un chemin $P_{s \to t}$ de la fronti\`ere $\mathscr{P}_\varepsilon$ qui consomme au plus $1/(1-\varepsilon)$ la consommation de $P'_{s \to t}$ sur chaque crit\`ere. En particulier, si $P'_{s \to t}$ est r\'ealisable, on d\'eduit que $P_{s \to t}$ est $1/(1-\varepsilon)$-r\'ealisable. 

\section{Synth\`ese}\label{sec-conc}

Le tableau \ref{tab-recap-comp} r\'esume les complexit\'es th\'eoriques des algorithmes de plus court chemin les plus connus ainsi que des algorithmes exacts ou approch\'es pour le PCCC vus dans ce rapport.
\begin{table}
\center

\caption{Diff\'erents algorithmes et leur complexit\'e}\label{tab-recap-comp}
%\hspace*{-14pt}
\begin{tabular}{|c|c|c|c|}
\hline
Nombre & Hypoth\`ese & Auteur & \\
de & sur & ou & Complexit\'e en temps\\
ressources & le graphe &  Nom de l'algorithme & \\
\hline
\multirow{2}*{$0$} & \multirow{2}*{$c_{ij} \in \mathbb{Q}$} & \multirow{2}*{Bellman\cite{Bellman1958}} & \multirow{2}*{$\mathcal{O}(mn)$}\\
& & & \\
\hline
\multirow{2}*{$0$} & $c_{ij} \in \mathbb{Q}$ & \multirow{2}*{Bellman} & \multirow{2}*{$\mathcal{O}(m+n)$}\\
 & G acyclique & & \\
\hline
\multirow{2}*{$0$} & \multirow{2}*{$c_{ij} \in \mathbb{Q}^+$} & \multirow{2}*{Dijkstra\cite{Dijkstra1959}} & \multirow{2}*{$\mathcal{O}((m+n)\, \log(n))$}\\
 & & & \\
\hline
\multirow{2}*{$0$} & \multirow{2}*{$c_{ij} \in \mathbb{Q}^+$} & \multirow{2}*{Dijkstra (tas de Fibonacci)} & \multirow{2}*{$\mathcal{O}(m+n\, \log(n))$}\\
& & & \\
\hline
\multirow{2}*{$1$} & \multirow{2}*{$c_{ij} \in \mathbb{Q}^+_*$} & \multirow{2}*{Programmation dynamique\cite{Dumitrescu2003}} & \multirow{2}*{$\mathcal{O}(m\, b_1)$}\\
& & & \\
\hline
\multirow{2}*{$R \in \mathbb{N}^+_*$} & \multirow{2}*{$c_{ij} \in \mathbb{Q}$} & \multirow{2}*{Programmation dynamique} & \multirow{2}*{$\mathcal{O}(e^n)$}\\
& & & \\
\hline
\multirow{2}*{$R \in \mathbb{N}^+_*$} & \multirow{2}*{$c_{ij} \in \mathbb{Q}$} & \multirow{2}*{$k$ plus courts chemins\cite{Eppstein1998}} & \multirow{2}*{$\mathcal{O}(m + Rn\, \log\ n +k), k = \mathcal{O}(e^n)$}\\
& & & \\
\hline
\hline
\multirow{3}*{1} & \multirow{3}*{$c_{ij} \in \mathbb{Q}^+$} & \multirow{3}*{FPTAS\cite{Phillips1993}} & \multirow{3}*{$\displaystyle \mathcal{O}\left(mn(1+\frac{1}{\varepsilon})+n^2(1+\frac{1}{\varepsilon})(\log(n)+\log(1+\frac{1}{\varepsilon}))\right)$}\\
& & & \\
& & & \\
\hline
\multirow{3}*{1} & \multirow{3}*{$c_{ij} \in \mathbb{Q}^+$} & \multirow{3}*{FPTAS multicrit\`ere\cite{Hansen1980}} & \multirow{3}*{$\displaystyle \mathcal{O}\left(m^2\frac{n^2}{\varepsilon}\log\left(\frac{n^2}{\varepsilon}\right)\right)$}\\
& & & \\
& & & \\
\hline
\multirow{3}*{1} & \multirow{3}*{$c_{ij} \in \mathbb{Q}^+$} & \multirow{3}*{FPTAS\cite{Hassin1992} \cite{Lorenz2001} \cite{Ergun2002}} & \multirow{3}*{$\displaystyle \mathcal{O}\left(\frac{mn}{\varepsilon}\right)$}\\
& & & \\
& & & \\
\hline
\multirow{3}*{$R \in \mathbb{N}^+_*$} & \multirow{3}*{$c_{ij} \in \mathbb{Q}^+$} & \multirow{2}*{Fronti\`ere Pareto $\varepsilon$-approch\'ee\cite{Papadimitriou2000}} & \multirow{3}*{$\displaystyle \mathcal{O}\left( \left(\frac{\log(MAJ)}{\varepsilon}\right)^R \frac{Rn^{4R+1}}{\varepsilon^{2R}}\right)$}\\
& & & \\
& & & \\
\hline
\end{tabular}
\end{table}

Le but de cette \'etude est d'\'etudier la possibilit\'e de g\'en\'erer des chemins contraints en temps ma\^{\i}tris\'e, dans l'optique d'acc\'el\'erer certains sch\'emas de g\'en\'eration de colonnes impliquant le PCCC comme sous-probl\`eme. Deux avantages pourraient \^etre tir\'es d'une telle d\'emarche~: d'une part, \^etre en mesure de g\'en\'erer une population diversifi\'ee de chemins lors de la phase d'initialisation~; d'autre part, acc\'el\'erer les premi\`eres it\'erations du sch\'ema. Les m\'ethodes les plus r\'epandues \`a ce jour pour r\'esoudre le PCCC, notamment dans ce cadre sp\'ecifique, sont celles de type programmation dynamique, qui offrent des solutions exactes, en temps et en espace exponentiel (ce sont encore ces m\^emes m\'ethodes qui ont inspir\'e les algorithmes approch\'es). Si nous avons \'evoqu\'e quelques m\'ethodes de simplification des instances (section~\ref{sec-reduc}), celles-ci ne garantissent pas pour autant la diminution de la complexit\'e th\'eorique. Quelle alternative avons-nous donc, quitte \`a s'affranchir de l'optimalit\'e, voire, de la r\'ealisabilit\'e~? 

La recherche de plus courts chemins dans le cadre sp\'ecifique du d\'eroulement d'un sch\'ema de g\'en\'eration de colonnes pose deux probl\`emes de nature diff\'erente. D'une part, celle de la difficult\'e intrins\`eque du probl\`eme contraint, \`a partir de 2 ressources~: si l'on souhaite rester dans un ordre polynomial de complexit\'e, il faut s'affranchir de l'exactitude en termes d'optimalit\'e et en termes de r\'ealisabilit\'e. D'autre part, les co\^uts manipul\'es sur les arcs ne sont plus les co\^uts initiaux de l'instance, mais des co\^uts r\'eduits issus de la r\'esolution du (PMR), qui peuvent \^etre n\'egatifs~: les hypoth\`eses faites sur la structure de co\^ut de l'instance initiale ne tiennent donc plus au cours des it\'erations. Or, ces hypoth\`eses interviennent conjointement dans les calculs de complexit\'e et des facteurs d'approximation. De sorte \`a exploiter les r\'esultats d'approximation, il faudrait donc \^etre en mesure de les \'etendre \`a un cadre plus g\'en\'eral (notamment, co\^uts rationnels n\'egatifs et positifs)~; en outre, il faudrait utiliser, voire, d\'efinir, un cadre pertinent pour l'approximation, l'approximation classique supposant des valeurs positives.

La diversification des chemins (notamment, dans la phase d'initialisation) sous-entend un certain balayage de l'espace des solutions. Parmi les approches que nous avons pr\'esent\'ees, la recherche des $k$ plus courts chemins d'une part (section~\ref{sub-resol-kpcc}), d'une fronti\`ere de Pareto $\varepsilon$-approch\'ee d'autre part (section~\ref{subsub-approx-multiobj-pcc}), vont dans le sens d'une telle exploration. Dans le premier cas, on cherche intensivement les meilleurs chemins contraints en terme de co\^ ut, tandis que dans le second cas, on balaye sporadiquement tout l'espace.

Pour un nombre de ressources $R$ donn\'e, l'algorithme d'Eppstein permet de trier en temps $\mathcal{O}(m + Rn\, \log(n))$ tous les chemins du graphe suivant le co\^ut (initial, ou r\'eduits dans le cadre de la g\'en\'eration de colonnes). Une fois le tas constitu\'e, il suffit pour trouver le chemin contraint optimal de d\'epiler les chemins jusqu'\`a trouver un chemin r\'ealisable (ou, dans le cadre d'un sch\'ema de g\'en\'eration de colonnes, quand le crit\`ere de co\^ut choisi pour la construction du tas est le co\^ut r\'eduit, l'optimalit\'e est prouv\'ee s'il n'existe pas de chemin r\'ealisable de co\^ut n\'egatif). Rien n'assure n\'eanmoins que les $k$ plus courts chemins, pour $k=\mathcal{O}(p(n))$ o\`u $p$ polyn\^ome, contiennent ne serait-ce qu'un chemin r\'ealisable pour les ressources (ni un chemin de co\^ut positif pour d\'emontrer l'optimalit\'e du sch\'ema). Par ailleurs, en triant sur le seul crit\`ere de co\^ut, on risque de g\'en\'erer des chemins peu diversifi\'es. Ainsi, en se limitant \`a chercher un nombre polynomial de chemins, on ne garantit pas l'obtention de chemins pertinents~; en prenant au contraire $k=\mathcal{O}(e^n)$, la construction du tas reste polynomiale en temps et en espace, et l'\'enum\'eration des chemins devient exponentielle en temps mais pas espace~: on n'obtient alors une alternative \`a la r\'esolution exacte par programmation dynamique qui est exponentielle en temps et en espace.

\`A partir de 2 ressources, l'obtention de solutions r\'ealisables n'\'etant plus garantie, l'approche multicrit\`ere semble prometteuse. 
%Si rien n'assure en th\'eorie que l'on g\'en\'erera des chemins r\'ealisables dans la fronti\`ere approch\'ee $\mathcal{P}_\varepsilon$, il faudrait \'evaluer en pratique le b\'en\'efice d'une telle d\'emarche (nombre de chemins r\'ealisables exhib\'es et diversification de ces chemins) mis en regard du co\^{u}t d'obtention de ces chemins (en fonction de $\varepsilon$). Cette \'etude devra tenir compte des caract\'eristiques de l'instance~: si les chemins r\'ealisables sont tous tr\`es proches des bornes par exemple, il sera plus difficile d'exhiber des chemins r\'ealisables~; si tous les chemins r\'ealisables ont un fort niveau de consommation pour chaque ressource, alors les chemins g\'en\'er\'es, r\'ealisables ou $1/(1-\varepsilon)$-r\'ealisables, seront vraisemblablement peu diversifi\'es. 
Bien plus que la recherche de la fronti\`ere de Pareto $\varepsilon$-approch\'ee elle-m\^{e}me, c'est le quadrillage de l'espace des valeurs et l'utilisation de la programmation dynamique sur les instances modifi\'ees qu'il faudra chercher \`a exploiter~: balayer l'espace des valeurs possibles et chercher une solution dans chaque hypercube semble \^etre une bonne strat\'egie en vue de la diversification des chemins d\`es l'initialisation. 
%-> Quadrillage ($\varepsilon$-r\'ealisable)~: on s'affranchit de l'exactitude du crit\`ere de co\^ut et/ou de l'exactitude des contraintes de ressource.

\bibliographystyle{alpha} 
\bibliography{Bibliographie}

\end{document}